\documentclass[final,a4paper]{scrartcl}
\pdfoutput=1 

\usepackage{amsmath,amsthm,mathtools,bm,relsize,xcolor,etoolbox,tikz,cmll,xifthen,enumerate,standalone,multicol,stackrel}
\usepackage{algpseudocode}
\usepackage{algorithm}
\usepackage[utf8]{inputenc} 
\usepackage[T1]{fontenc}
\usepackage[bitstream-charter,cal=cmcal]{mathdesign}

\usepackage[english]{babel}
\usepackage{csquotes}

\AtBeginDocument{
  \usepackage[
    activate={true,nocompatibility}
  , final
  , tracking=true
  , kerning=true
  , spacing=true
  , kerning=true
  , spacing=true
  , babel=true
]{microtype}}

\let\Gamma\varGamma
\let\Delta\varDelta
\let\Lambda\varLambda
\let\Xi\varXi
\let\Pi\varPi
\let\Sigma\varSigma
\let\Upsilon\varUpsilon
\let\Omega\varOmega

\theoremstyle{plain}
\newtheorem{theorem}{Theorem}[section]
\newtheorem{lemma}{Lemma}[section]

\newtheorem{proposition}{Proposition}[section]

\newtheorem{observation}{Observation}[section]
\theoremstyle{definition}

\theoremstyle{remark}

\newtheorem{example}{Example}[section]

\theoremstyle{plain}

\usetikzlibrary{arrows,positioning,shapes,trees,topaths,chains,calc,decorations.pathreplacing,backgrounds}

\tikzset{
  treefig/.style={
    level distance=20pt
  , text height=1.5ex
  , text depth=0.75ex
  , sibling distance=20pt
  , inner sep=1.5pt
  }
}

\renewcommand{\phi}{\varphi}
\renewcommand{\epsilon}{\varepsilon}
\renewcommand{\theta}{\vartheta}
\newcommand{\Nat}{\mathbb{N}}

\newcommand{\lang}{\mathcal{L}}
\newcommand{\csub}{\multimapinv}
\newcommand{\T}{\mathrm{T}}
\renewcommand{\S}{\mathrm{S}}
\newcommand{\TT}[3]{\T(#1)^{#2}_{#3}}
\newcommand{\lTT}[3]{\widetilde{\T}(#1)^{#2}_{#3}}
\newcommand{\sTT}[3]{\S(#1)^{#2}_{#3}}
\newcommand{\sTheta}[2]{\widehat\Theta^{#1}_{#2}}
\newcommand{\lsTT}[3]{\widetilde{\S}(#1)^{#2}_{#3}}
\newcommand{\TSigma}{\T_{\Sigma}}
\newcommand{\TDelta}{\T_{\Delta}}

\newcommand{\Id}[1]{\textnormal{Id}_{#1}}

\newcommand{\card}[1]{\lvert #1 \rvert}
\newcommand{\abs}[1]{\lvert #1 \rvert}

\newcommand{\stepOI}{%
  \mathrel{\vbox{\offinterlineskip\ialign{%
    \hfil##\hfil\cr
    \fontsize{5}{0}\selectfont\text{\!\!\textnormal{\textbf{OI}}}\cr
    \vspace{.5pt}\cr
    \(\Rightarrow\)\cr
}}}}

\newcommand{\derivOI}[2]{\mathrel{\mathrlap{\stepOI}\phantom{\Rightarrow}_{#1}%
    \if\relax\detokenize{#2}\relax\relax\else^{#2}\fi}}

\DeclareMathOperator{\rk}{rk}

\DeclareMathOperator{\pos}{pos}

\DeclareMathOperator{\lin}{lin}

\newcommand{\sepstars}{\bigskip\par\centerline{*\,*\,*}\medskip\par}%

\usepackage[textsize=footnotesize,linecolor=yellow,backgroundcolor=yellow]{todonotes}
\presetkeys{todonotes}{noline}{}

\usepackage[pdfusetitle,hidelinks]{hyperref}

\addtokomafont{disposition}{\rmfamily}

\title{Linear Context-Free Tree Languages and Inverse Homomorphisms}
\author{Johannes Osterholzer \qquad Toni Dietze \qquad Luisa Herrmann
  \\[3mm]\smaller Faculty of Computer Science, Technische Universität Dresden
}

\begin{document}
\maketitle

\begin{abstract}
We prove that the class of linear context-free tree languages is not closed under inverse linear tree homomorphisms.  The proof is by contradiction: we encode Dyck words into a context-free tree language and prove that its preimage under a certain linear tree homomorphism cannot be generated by any context-free tree grammar.  A positive result can still be obtained: the linear monadic context-free tree languages \emph{are} closed under inverse linear tree homomorphisms.
\end{abstract}

\section{Introduction}
Context-free tree grammars (cftg), introduced by Rounds \cite{Rounds1970}, generalize the concept of context-free rewriting to the realm of tree languages.  A sentential form of a cftg is a tree labeled by terminal and nonterminal symbols.  In contrast to a regular tree grammar, any node of a sentential form may be labeled by a nonterminal, not just a leaf node.  
In general, cftg can \emph{copy} parts of a sentential form in the application of a production, and a lot of their complexity in comparison to regular tree grammars is due to the interplay between copying and nondeterminism (cf.\ e.g.
\cite{Engelfriet1977}).

While Rounds viewed context-free tree grammars as a promising model for mathematical linguistics,
the main motivation for studying cftg in the 1970s and 80s was their application in the theory of recursive program schemes \cite{Arnold1979,Engelfriet1977,Nivat1974}.
Evidently, in this context, the ability of cftg to copy is essential -- after all, it is quite a harsh restriction on a program to demand that each formal parameter of a function is used at most once in its body.

In the recent years, there has been renewed interest in cftg in the area of syntax-based natural language processing \cite{Kanazawa2013,Kepser2006,Kepser2011,Maletti2012,Nederhof2012}, where tree languages are used to express the linguistic structure of the processed sentences.  Here, cftg allow modelling particular linguistic phenomena, which are described as \emph{mildly context-sensitive}.  In contrast to recursive program schemes, in this area only non-copying, or \emph{linear}, cftg (l-cftg) are considered, as there is no linguistic motivation for copying, and as the copying power of cftg makes their membership problem computationally hard \cite{Osterholzer2015,Rounds1973}.

The modular design of syntax-based language processing systems requires that the utilized class of tree languages \(\mathcal{C}\) possesses a number of closure properties.  In particular, for translation tasks it is important that \(\mathcal{C}\) is closed under application of linear extended tree transducers (l-xtt).  This transducer model was first described by Rounds \cite{Rounds1970} (under another name), and further investigated, i.a., in \cite{Arnold1982,Fulop2011,Maletti2009a}.  Unfortunately, the closure under l-xtt does \emph{not} hold when \(\mathcal{C}\) is the class of context-free tree languages.  This is due to a theorem of Arnold and Dauchet, who proved that the context-free tree languages are not closed under inverse linear tree homomorphisms~\cite{Arnold1978}.  Trivially, every inverse linear tree homomorphism can be computed by an l-xtt.
The proof in~\cite{Arnold1978} works by constructing a \emph{copying} cftg \(G\), and the preimage of the tree language of \(G\) under a certain tree homomorphism is shown to be non-context-free.  

But since copying is not required anyway -- are maybe the \emph{linear} context-free tree languages closed under inverse linear tree homomorphisms?  In this work, we answer this question in the negative: there are an l-cftg \(G_{\mathrm{ex}}\) and a linear tree homomorphism \(h\) such that \(L=h^{-1}(\lang(G_{\mathrm{ex}}))\) is \emph{not} a context-free tree language.\footnote{Where, of course, \(\lang(G_{\mathrm{ex}})\) is the tree language of \(G_{\mathrm{ex}}\).}

The intuition behind our proof is as follows.  Every tree \(t\) in \(L\) is of the form
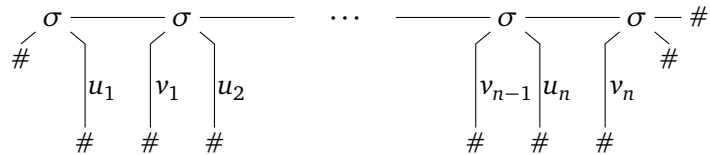
\begin{center}
  
\begin{tikzpicture}[node distance=7ex, text height=1.5ex, text depth=0.25ex, every node/.style={inner sep=1pt,minimum width=2.5ex,minimum height=2.5ex}]

  \begin{scope}[every on chain/.style={join}, start chain=1]
    \node [on chain=1] (s1) {\(\sigma\)};
    \node [on chain=1] (s2) {\(\sigma\)};
    \node [on chain=1] {\(\hphantom{M}\cdots\hphantom{M}\)};
    \node [on chain=1] (s3) {\(\sigma\)};
    \node [on chain=1] (s4) {\(\sigma\)};
    \node [on chain=1,xshift=-5ex] {\(\#\)};
  \end{scope}

  \draw (s1) -- ++(-12pt,-10pt) node[yshift=-6pt] {\(\#\)};
  \draw (s1) -- ++( 12pt,-10pt) -- ++(0,-3em) node[midway,right] {\(u_1\)} node[yshift=-6pt] {\(\#\)};
  \draw (s2) -- ++(-12pt,-10pt) -- ++(0,-3em) node[midway,right] {\(v_1\)} node[yshift=-6pt] {\(\#\)};
  \draw (s2) -- ++( 12pt,-10pt) -- ++(0,-3em) node[midway,right] {\(u_2\)} node[yshift=-6pt] {\(\#\)};
  \draw (s3) -- ++(-12pt,-10pt) -- ++(0,-3em) node[midway,right] {\(v_{n-1}\)} node[yshift=-6pt] {\(\#\)};
  \draw (s3) -- ++( 12pt,-10pt) -- ++(0,-3em) node[midway,right] {\(u_n\)} node[yshift=-6pt] {\(\#\)};
  \draw (s4) -- ++(-12pt,-10pt) -- ++(0,-3em) node[midway,right] {\(v_n\)} node[yshift=-6pt] {\(\#\)};
  \draw (s4) -- ++( 12pt,-10pt) node[yshift=-6pt] {\(\#\)};

\end{tikzpicture}

\end{center}
for some \(n \geq 1\) and monadic trees \(u_1\), \(v_1\), \ldots, \(u_n\), \(v_n\).  Here, the root of \(t\) is the leftmost symbol~\(\sigma\).  The subtrees \(u_i\), \(v_i\), called \emph{chains} in the following, are built up over a parenthesis alphabet, such that the chains \(u_i\) contain only opening parentheses, the chains \(v_i\) only closing parentheses, and \(u_1^R v_1 \cdots u_n^R v_n\) is a well-parenthesized word.\footnote{Here, \(w^R\) denotes the reversal of the word \(w\).}

If one were to cut such a tree \(t\) into two parts \(t_1\) and \(t_2\), right through an edge between two \(\sigma\)s, then one could observe that there are some chains \(u_j\) in \(t_1\) which contain opening parentheses which are not closed in \(t_1\), but only in \(t_2\).  A similar observation holds of course for some chains \(v_j\) in \(t_2\).  These chains \(u_j\) and \(v_j\) will be called \emph{critical chains}, and their ``unclosed'' parts \emph{defects.}

We assume that there is some (not necessarily linear) cftg \(G\) with \(\lang(G) = L\), and show that if \(G\) exists, then it can be assumed to be of a special normal form.  We analyze the derivations of such a \(G\) in normal form.  A derivation of a tree \(t\) as above begins with a subderivation
\[A(\#, \ldots, \#) \Rightarrow_G^* B(s_1, \ldots, s_p, \#)\,\text{,}\]
where \(A\) and \(B\) are nonterminals of \(G\), and \(s_1\), \ldots, \(s_p\) are chains over the parenthesis alphabet.  After that, the derivation continues with
\[B(s_1, \ldots, s_p, \#) \Rightarrow_G C(s'_1, \ldots, s'_p, D(s'_{p+1}, \ldots, s'_{2p}, \#))\,\text{,}\]
for some nonterminals \(C\) and \(D\) and \(s'_1\), \ldots, \(s'_{2p} \in \{s_1, \ldots, s_p\}\).  Finally, \(C\) and \(D\) derive some terminal trees \(t_1\) and \(t_2\), respectively.  So a derivation of \(t\) in \(G\) ``cuts'' \(t\) into two pieces as described above!

If \(G\) exists, it must therefore prepare the defects of \(t_1\) and \(t_2\) such that they ``fit together'', and it can only do so in the initial subderivation \(A(\#, \ldots, \#) \Rightarrow_G^* B(s_1, \ldots, s_p, \#)\).  But there are only finitely many arguments of \(A\) in which the defects could be prepared.
We give a sequence of trees in \(L\) such that the number of their defects is strictly increasing, no matter how they are cut apart.  Then there is some tree \(t\) in this sequence whose defects cannot be prepared fully.  Hence it is possible to show by a pumping argument that if \(t \in \lang(G)\), then there is also a tree \(t' \in \lang(G)\) whose respective parts do not fit together, and therefore \(t' \notin L\).
Thus the existence of \(G\) is ruled out.

We conclude our work with a positive result: the tree languages of linear \emph{monadic} cftg (lm-cftg), i.e.\ of l-cftg where each nonterminal has at most one successor, \emph{are} closed under inverse linear tree homomorphisms.  The importance of lm-cftg is underscored by their expressive equivalence to the well-known linguistic formalism of \emph{tree-adjoining grammars} \cite{Kepser2011,Osterholzer2015b}.
Our proof is based on the Greibach normal form of lm-cftg \cite{Fujiyoshi2006}.  In fact, the closure of Greibach cftg under inverse linear tree homomorphisms was already proven by Arnold and Leguy \cite{Arnold1980}, but their construction results in a copying cftg of higher nonterminal arity.

The article is organized as follows.
After establishing some preliminaries in Section~\ref{sec:preliminaries}, we define the tree language \(L\) in Section~\ref{sec:language-L}.
In Section~\ref{sec:grammar}, the grammar \(G_{\mathrm{ex}}\) is introduced, while Section~\ref{sec:homomorphism-preimage} contains the definition of the homomorphism \(h\) and some easy observations on \(L\).  In Section~\ref{sec:normal-form} we work out a normal form for the assumed cftg \(G\), which allows us to define the concept of \emph{derivation trees} of \(G\) in Section~\ref{sec:derivation-trees}.  This concept facilitates the analysis of the derivations in \(G\).  Section~\ref{sec:dyck-words-sequences} contains some properties about factorizations of Dyck words, which formalize the idea of cutting \(t\) into two.  Finally, in Section~\ref{sec:witness-langg-ne} we give a counterexample, and rule out the existence of \(G\).  Section~\ref{sec:lmcftg} is about the positive result for lm-cftg.

\section{Preliminaries}
\label{sec:preliminaries}
The set of natural numbers with zero is denoted by \(\Nat\).  For every \(m\), \(n \in \Nat\), the set \(\{i \in \Nat \mid m \leq i \leq n\}\) is denoted by \([m,n]\), and the set \([1,n]\) by \([n]\).

An alphabet is a finite nonempty set.  The set of words over \(A\) is \(A^*\), the empty word is \(\epsilon\), and \(A^+ = A^* \setminus \{\epsilon\}\).  We often abbreviate \(\{w\}^*\) by \(w^*\), and \(\{w\}^+\) by \(w^+\), for \(w \in A^*\).
Let \(w = a_1 \cdots a_n\) with \(a_1\), \ldots, \(a_n \in A\) for some \(n \in \Nat\).  Then \(\abs{w} = n\), and \(w^R = a_n \cdots a_1\), the reversal of \(w\).
Moreover, let \(v \in A^*\).  We say that \(v\) is a \emph{factor} of \(w\) if there are \(w'\), \(w'' \in A^*\) such that \(w = w' v w''\).  If, additionally, \(w' = \epsilon\) (resp.\ \(w'' = \epsilon\)), then \(v\) is a \emph{prefix} or \emph{left factor} (resp.\ \emph{suffix} or \emph{right factor}) of \(w\).

Let \(A\) be an alphabet such that \(A = B \cup C\), \(B \cap C = \emptyset\), and there is a one-to-one relation between the elements of \(B\) and \(C\).  Define the \emph{Dyck congruence} \(\equiv\) to be the smallest congruence relation on \(A^*\) such that \(b c \equiv \epsilon\) for each pair of related elements \(b \in B\) and \(c \in C\).  We also say that \(c\) \emph{acts as right inverse} to \(b\).  The \emph{Dyck language} over \(A\) is the set \(D_A^* = \{w \in A^* \mid w \equiv \epsilon\}\).  By saying that \(w \in A^*\) \emph{reduces to} \(v\) (resp.\ \(v\) is the \emph{reduction} of \(w\)), we mean that \(v\) is the (unique) shortest word in \(A^*\) such that \(v \equiv w\).

\paragraph{Trees} An alphabet \(\Sigma\) equipped with a function \(\rk_\Sigma \colon \Sigma \to \Nat\) is a \emph{ranked alphabet}.  Let \(\Sigma\) be a ranked alphabet.  When \(\Sigma\) is obvious, we write \(\rk\) instead of \(\rk_\Sigma\).  Let \(k \in \Nat\).  Then \(\Sigma^{(k)} = \rk^{-1}(k)\).  We often write \(\sigma^{(k)}\) and mean that \(\rk(\sigma) = k\).

Let \(U\) be a set and \(\Lambda\) denote \(\Sigma \cup U \cup C\), where \(C\) consists of the three symbols `(', `)', and~`,'.
The set \(\TSigma(U)\) of \emph{trees (over \(\Sigma\) indexed by \(U\))} is the smallest set \(T \subseteq \Lambda^*\) such that \(U \subseteq T\), and 
for every \(k \in \Nat\), \(\sigma \in \Sigma^{(k)}\), and \(t_1\), \ldots, \(t_k \in T\), we also have that \(\sigma(t_1, \ldots, t_k) \in T\).
A tree \(\alpha()\), \(\alpha \in \Sigma^{(0)}\), is abbreviated by \(\alpha\), a tree \(\gamma(t)\), \(\gamma \in \Sigma^{(1)}\), by \(\gamma t\), and \(\TSigma(\emptyset)\) by \(\TSigma\).  The notation \(\gamma t\) suggests a bijection between \(\Sigma^* U\) and \(\TSigma(U)\) for monadic ranked alphabets \(\Sigma\) (i.e.\ \(\Sigma = \Sigma^{(1)}\)), and in fact we will often confuse such monadic trees with words in writing.

Let \(s\), \(t \in \TSigma(U)\).  The set of \emph{positions (Gorn addresses) of \(t\)} is denoted by \(\pos(t) \subseteq \Nat^*\).  The number of occurrences of a symbol \(\sigma \in \Sigma\) in \(t\) is written \(\abs{t}_\sigma\).  The \emph{size} of \(t\) is \(\abs{t} = \sum_{\sigma\in\Sigma}\abs{t}_\sigma\).  Denote the \emph{label} of \(t\) at its position \(w\) by \(t(w)\), and the \emph{subtree} of \(t\) at \(w\) by \(t\vert_w\).  The result of \emph{replacing} the subtree \(t\vert_w\) in \(t\) by \(s\) is \(t[s]_w\).
Fix the infinite set of \emph{variables} \(X = \{x_1,x_2,\ldots\}\).  For each \(k \in \Nat\), let \(X_k = \{x_i \mid i \in [k]\}\).  Given \(n\), \(k \in \Nat\), \(t \in \TSigma(X_n)\), and \(s_1\), \ldots, \(s_n \in \TSigma(X_k)\), denote by \(t[s_1,\ldots, s_n]\) the result of substituting \(s_i\) for each occurrence of \(x_i\) in \(t\), where \(i \in [n]\).  Sometimes, especially when no other variable is used, we will write \(x\) instead of \(x_1\).

\paragraph{Magmoids} We will heavily use the notation introduced with the concept of \emph{magmoids}  \cite{Arnold1978a,Arnold1982}.  Let \(k\), \(n \in \Nat\).  Then the set \(\{\langle k, t_1, \ldots, t_n \rangle \mid t_1, \ldots, t_n \in \TSigma(X_k)\}\) is denoted by \(\TT{\Sigma}{n}{k}\).
It is customary to omit the component \(k\) from such a tuple, and we will do so from now on.  

Let \(\TT{\Sigma}{}{} = \bigcup_{n, k \in \Nat}\TT{\Sigma}{n}{k}\).  We identify the sets \(\TSigma(X_k)\) and \(\TT{\Sigma}{1}{k}\) and write \(t\) instead of~\(\langle t \rangle\).  Moreover, we follow the convention of identifying the tree \(\sigma(x_1, \ldots, x_k) \in \TT{\Sigma}{1}{k}\) with the symbol \(\sigma \in \Sigma^{(k)}\).

The set of all \(u \in \TT{\Sigma}{n}{k}\) such that the left-to-right sequence of variables in \(u\) is  \(x_1\), \ldots, \(x_k\), is denoted by \(\lTT{\Sigma}{n}{k}\).  
The set \(\Theta^n_k\) of \emph{torsions} is \(\{\langle x_{i_1}, \ldots, x_{i_n}\rangle \mid i_1, \ldots, i_n \in [k]\}\).  Note that \(\Theta^n_k \subseteq \TT{\Sigma}{n}{k}\).  We may also understand a torsion \(\theta \in \Theta^n_k\) as a function \(\theta \colon [n] \to [k]\), such that \(\theta(i) = j\) if and only if \(\theta_i = x_j\).
The torsion \(\langle x_1, \ldots, x_n\rangle \in \Theta^n_n\) is denoted by \(\Id{n}\), and the torsion \(\langle x_i \rangle \in \TT{\Sigma}{1}{k}\), \(i \in [k]\), by \(\pi^k_i\) (when \(k\) is clear from the context, we write \(\pi_i\) instead).  Then \(\pi_i \cdot u\) is the \(i\)-th component of the tuple \(u\).
A tuple \(u \in \lTT{\Sigma}{n}{k}\) is said to be \emph{torsion-free}.  We write \(\lin(u)\) for the (unique) tuple \((v, \theta) \in \lTT{\Sigma}{n}{m} \times \Theta^m_k\) such that \(u = v \cdot \theta\).  

Let \(n\), \(\ell\), \(k \in \Nat\), and let \(u \in \TT{\Sigma}{n}{\ell}\), \(v \in \TT{\Sigma}{\ell}{k}\).  We define \(u \cdot v \in \TT{\Sigma}{n}{k}\) by
\[u \cdot v = \bigl\langle u_1[v_1, \ldots, v_\ell], \ldots, u_n[v_1, \ldots, v_\ell] \bigr\rangle\,\text{.}\]
Note that the operation \(\cdot\) is associative \cite[Prop.~2.4]{Goguen1977}.  If \(u \in \TT{\Sigma}{n}{n}\), then let \(u^{0} = \Id{n}\) and \(u^{(j+1)} = u \cdot u^{j}\) for every \(j \geq 0\).
Moreover, given \(u \in \TT{\Sigma}{n}{k}\) and \(v \in \TT{\Sigma}{\ell}{k}\), define \([u, v] \in \TT{\Sigma}{n+\ell}{k}\) by
\[[u,v] = \langle u_1, \ldots, u_n, v_1, \ldots, v_\ell \rangle\,\text{.}\]
Clearly, this operation is associative, so we will write, e.g., \([u,v,t]\) instead of \([[u,v], t]\).

\paragraph{Trees with a spine}
We introduce the following special notation which will be helpful to denote the trees we deal with.
Let \(n\), \(k \in \Nat\).  Then
\[\sTT{\Sigma}{n}{k} = \bigl\{[u,x_{k+1}] \bigm\vert u \in \TT{\Sigma}{n}{k}\bigr\}\,\text{,} \quad \lsTT{\Sigma}{n}{k} = \lTT{\Sigma}{n+1}{k+1} \cap \sTT{\Sigma}{n}{k}\,\text{,} \quad \text{and} \quad \sTheta{n}{k} = \Theta^{n+1}_{k+1} \cap \sTT{\Sigma}{n}{k}\,\text{.}\]
Moreover, for every \(s \in \TT{\Sigma}{1}{n+1}\) and \(t \in \TT{\Sigma}{1}{k}\), let
\(s \csub t = s \cdot [\Id{n},\, t]\).
In an expression containing \(\cdot\) and \(\csub\), we assume \(\cdot\) to bind stronger than \(\csub\).

\paragraph{Context-free tree grammars}
A \emph{context-free tree grammar (cftg)} over \(\Sigma\) is a tuple \(G = (N, \Sigma, \eta_0, P)\) such that \(\Sigma\) and \(N\) are disjoint ranked alphabets (of \emph{terminal} resp.\ \emph{nonterminal symbols}), \(\eta_0 \in \TT{N \cup \Sigma}{1}{0}\) (the \emph{axiom}\footnote{Just like in the word case, the use of an axiom instead of an initial nonterminal has no impact on the generative power of cftg.  But it will be technically convenient to use an axiom.}) and \(P\) is a finite set of \emph{productions} of the form
\(A(x_1, \ldots, x_k) \to t\) for some \(k \in \Nat\), \(A \in N^{(k)}\), and \(t \in \TT{N \cup \Sigma}{1}{k}\).  Let \(G = (N, \Sigma, S, P)\) be a cftg, and let \(n\), \(m \in \Nat\).

The cftg \(G\) is said to be \emph{linear} (resp.\ \emph{nondeleting}) if in every production \(A(x_1, \ldots, x_k) \to t\) in \(P\) the right-hand side \(t\) contains each variable \(x_i\), \(i \in [k]\), at least (resp.\ at most) once.  A linear and nondeleting cftg is \emph{simple}.  The cftg \(G\) is a \emph{regular tree grammar (rtg)} if \(N = N^{(0)}\).  Finally, \(G\) is \emph{monadic} if \(N = N^{(0)} \cup N^{(1)}\).  Linear (and monadic) cftg are abbreviated \emph{l-cftg} (\emph{lm-cftg}).

Given \(\eta\), \(\zeta \in \TT{N \cup \Sigma}{1}{m}\), we write \(\eta \Rightarrow_G \zeta\) if there are \(A(x_1, \ldots, x_k) \to t\) in \(P\) and \(\kappa \in \TT{N \cup \Sigma}{1}{m+1}\), \(\tau \in \TT{N \cup \Sigma}{k}{m}\) such that \(\kappa\) contains \(x_{m+1}\) exactly once,
\begin{equation}
  \eta = \kappa \cdot [\Id{m},\, A \cdot \tau]\,\text{,} \qquad \text{and} \qquad \zeta = \kappa \cdot [\Id{m},\, t \cdot \tau]\,\text{.}\label{eq:derivation}
\end{equation}

If \(\eta\), \(\zeta \in \TT{N \cup \Sigma}{n}{m}\) instead, for \(n>1\), write \(\eta \Rightarrow_G \zeta\) if there is some \(i \in [n]\) such that \(\pi_i \cdot \eta \Rightarrow_G \pi_i \cdot \zeta\), and \(\pi_j \cdot \eta = \pi_j \cdot \zeta\) for every \(j \in [n]\), \(j \ne i\).
Let \(k \in \Nat\) and \(\eta \in \TT{N \cup \Sigma}{1}{k}\).  Then the set \(\{t \in \TT{\Sigma}{1}{k} \mid \eta \Rightarrow_G^* t\}\) is denoted by \(\lang(G, \eta)\), and
the \emph{tree language of \(G\)}, denoted by \(\lang(G)\), is \(\lang(G,\eta_0)\).  We call \(L \subseteq \TT{\Sigma}{1}{0}\) a \emph{(linear) (monadic) context-free tree language} if there is a (linear) (monadic) cftg \(G\) with \(L = \lang(G)\).  Two cftg \(G\) and \(G'\) are \emph{equivalent} if \(\lang(G) = \lang(G')\).

Recall that there are two restricted modes of derivation for cftg, the OI and the IO mode.  In a nutshell, the OI mode requires that a nonterminal may only be rewritten if it occurs outermost in a sentential form.
Formally, we write \(\eta \derivOI{G}{} \zeta\) if, additionally to the conditions of~\eqref{eq:derivation}, the path from the root of \(\kappa\) to the single occurence of \(x_{m+1}\) is only labelled by symbols from \(\Sigma \cup X\).  The relation \(\derivOI{G}{}\) is extended to \(\TT{N \cup \Sigma}{n}{m}\) in the same manner as above.  

Dually, in the IO mode, a nonterminal may only be rewritten if it occurs innermost.  
It is well-known that every unrestricted derivation can be emulated by one that is OI, and therefore \(\lang(G) = \{t \in \TT{\Sigma}{1}{0} \mid \eta_0 \derivOI{G}{*} t\}\).  Under the IO mode there may indeed be some trees in \(\lang(G)\) which cannot be derived in this restricted manner \cite{Engelfriet1977,Fischer1968a}.

The following lemma fulfills the role of a basic technical lemma on context-free word grammars (e.g.\ \cite[Lemma~3.3.1]{Harrison1978}).  As we must count the number of derivation steps, OI derivations are used.
\begin{lemma}[{\cite[Lemma~2]{Arnold1980}}]
  \label{lem:technical}
  Let \(G = (N, \Sigma, \eta_0, P)\) be a cftg, \(n\), \(p\), \(q\), \(r \in \Nat\), \(\eta \in \TT{N \cup \Sigma}{p}{q}\), \(\kappa \in \TT{N \cup \Sigma}{q}{r}\), and \(t \in \TT{\Sigma}{p}{r}\).
  Then
  \(\eta \cdot \kappa \derivOI{G}{n} t\)
  if and only if there are \(k\), \(m\), \(\ell \in \Nat\), \(\tilde u \in \lTT{\Sigma}{p}{\ell}\), \(\theta \in \Theta^{\ell}_{q}\), and \(v \in \TT{\Sigma}{\ell}{r}\) such that
  \[t = \tilde u \cdot v\,\text{,}
    \qquad
    \eta \derivOI{G}{k} \tilde u \cdot \theta\,\text{,}
    \qquad
    \theta \cdot \kappa \derivOI{G}{m} v\,\text{,}
    \qquad \text{and} \qquad
    k + m = n\,\text{.}
  \]
\end{lemma}

A cftg \(G\) is said to be \emph{total} if \(\lang(G,A) \ne \emptyset\) for every nonterminal \(A\) of \(G\).  As the following lemma shows, we may always assume that a cftg is total.

\begin{lemma}[{\cite[Appendix]{Arnold1976a}}]
  \label{lem:cftg-total}
  For every cftg \(G\), an equivalent total cftg \(G'\) can be constructed.
\end{lemma}
The proof in \cite{Arnold1976a} assumes that \(G\) is in normal form, but with an evident generalization it also goes through without this assumption.  The proof's idea is to introduce the production \(A \to \#\), where \(\#\) is some dummy symbol, for every non-productive nonterminal \(A\) of \(G\), i.e.\ with \(\lang(G,A) = \emptyset\).
Of course, care must be taken that this dummy symbol is not produced in the course of a derivation in \(G'\) which was blocked before in \(G\).
Therefore every nonterminal \(A \in N^{(k)}\) is annotated with a set \(\alpha \subseteq [k]\) of forbidden indices, which prevents choosing a non-productive nonterminal.  Apart from this annotation, the construction does not alter the shape of the productions of \(G\).

\paragraph{Tree Homomorphisms}  Let \(\Sigma\) and \(\Delta\) be ranked alphabets.  A mapping \(h \colon \Sigma \to \Delta\) is said to be a \emph{tree homomorphism} if \(h\bigl(\Sigma^{(k)}\bigr) \subseteq \TDelta(X_k)\) for every \(k \in \Nat\).  We extend \(h\) to a mapping \(\widehat h \colon \TSigma(X) \to \TDelta(X)\) by setting \(\widehat h(x_i) = x_i\) for every \(i \in \Nat\) and
\[\widehat h(\sigma(t_1, \ldots, t_k)) = h(\sigma)[\widehat h(t_1), \ldots, \widehat h(t_k)]\]
for every \(k \in \Nat\), \(\sigma \in \Sigma^{(k)}\), and \(t_1\), \ldots, \(t_k \in \TSigma(X)\).  In the following, we will no longer distinguish between \(h\) and \(\widehat h\).

We recall the following properties of tree homomorphisms (cf.\ \cite{Arnold1980}).  Let \(h \colon \TSigma(X) \to \TDelta(X)\) be a tree homomorphism, and for every \(\sigma \in \Sigma^{(k)}\), \(k \in \Nat\), let \(h(\sigma) = \tilde t_\sigma \cdot \theta_\sigma\), where \(\tilde t_\sigma \in \lTT{\Delta}{1}{\ell}\) and \(\theta_\sigma \in \Theta^\ell_k\), for an \(\ell \in \Nat\).  We say that \(h\) is
\emph{linear} (resp.\ \emph{nondeleting}) if \(\theta_\sigma\) is injective (resp.\ surjective),
and \emph{alphabetic} or a \emph{delabeling (d\'emarquage)} if \(\tilde t_\sigma \in \Sigma \cup X\),
for every \(\sigma \in \Sigma\).
Moreover, \(h\) is \emph{simple} if it is linear and nondeleting. 
Lastly, \(h\) is \emph{elementary ordered (\'el\'ementaire ordonn\'e)} if there are \(\sigma \in \Sigma^{(n)}\), \(\delta_1 \in \Delta^{(n-k+1)}\), \(\delta_2 \in \Delta^{(k)}\), \(n\), \(k \in \Nat\), and \(\ell \in [n+1]\) such that
\[h(\sigma) = \delta_1\bigl(x_1, \ldots, x_{\ell-1}, \delta_2(x_\ell, \ldots, x_{\ell+k-1}), x_{\ell+k}, \ldots, x_n\bigr)\]
and \(h(\omega) = \omega\) for every \(\omega \in \Sigma \setminus \{\sigma\}\).

\section{\boldmath The tree language \(L\)}
\label{sec:language-L}
We start out by introducing the cftg \(G_{\mathrm{ex}}\).  The preimage \(L\) of \(\lang(G_{\mathrm{ex}})\) under a simple tree homomorphism \(h\), introduced afterwards, will be shown to be non-context-free later on.

\subsection{\boldmath The grammar \(G_{\mathrm{ex}}\)}

\label{sec:grammar}
Let \(\Delta = \{\delta_1^{(2)}, \delta_2^{(2)}, \#^{(0)}\} \cup \Gamma\), where \(\Gamma = \{a^{(1)}, b^{(1)}, c^{(1)}, d^{(1)}\}\).
Consider the simple cftg \(G_{\mathrm{ex}} = (N_{\mathrm{ex}}, \Delta, \eta_{\mathrm{ex}}, P_{\mathrm{ex}})\) with nonterminal set \(N_{\mathrm{ex}} = \{A^{(3)}\}\), axiom \(\eta_{\mathrm{ex}} = \delta_1\bigl(\#, A(c \#, d\#, \delta_2(\#,\#))\bigr)\), and productions in \(P_{\mathrm{ex}}\) given by
\begin{align*}
  A(x_1,x_2,x_3) \to {} & A(a x_1, b x_2, x_3) \\
  {} + {}&A\bigl(c c x_1, d \#,\, A(c \#, d d x_2, x_3)\bigr) \\
  {} + {}&\delta_2\bigl(c x_1,\delta_1(d x_2,x_3)\bigr)\,\text{.}
\end{align*}

\begin{example}
  \label{ex:derivation}
  The following is an example derivation of a tree in \(\lang(G_{\mathrm{ex}})\).
  \begin{alignat*}{2}
    \eta_{\mathrm{ex}} & {}={} &&\delta_1(\#,x) \csub A(c \#, d\#, x) \csub \delta_2(\#,\#) \\
    &{} \Rightarrow_{G_{\mathrm{ex}}}^* {} &&\delta_1(\#,x) \csub A(a^2c \#, b^2d\#, x) \csub \delta_2(\#,\#)\\
    &{} \Rightarrow_{G_{\mathrm{ex}}} {} &&\delta_1(\#,x) \csub A(c^2a^2c \#, d\#, x) \csub A(c\#, d^2b^2d\#, x) \csub \delta_2(\#,\#)\\
    &{}\Rightarrow_{G_{\mathrm{ex}}}^* {} &&\delta_1(\#,x) \csub A(ac^2a^2c \#, bd\#, x) \csub A(a^2c\#, b^2d^2b^2d\#, x) \csub \delta_2(\#,\#)\\
    &{}\Rightarrow_{G_{\mathrm{ex}}}^*{} &&\delta_1(\#,x) \csub \delta_2(cac^2a^2c \#, x) \csub \delta_1(dbd\#, x) \\
    &&&\hphantom{\delta_1(\#,x)} {} \csub \delta_2(ca^2c\#, x) \csub \delta_1(db^2d^2b^2d\#, x) \csub \delta_2(\#,\#)\,\text{.}
  \end{alignat*}
\end{example}

\subsection{\boldmath The homomorphism \(h\) and its preimage}
\label{sec:homomorphism-preimage}
Let \(\Sigma = \{\sigma^{(3)}, \#^{(0)}\} \cup \Gamma\) and let \(h \colon \TSigma(X) \to \TDelta(X)\) be the simple tree homomorphism such that
\[h(\sigma(x_1,x_2,x_3)) = \delta_1(x_1,\delta_2(x_2,x_3)) \quad \text{and} \quad h(\omega) = \omega \text{ for each } \omega \in \Sigma \setminus \{\sigma\}\,\text{.}\]
In the following, we will analyse the tree language \(L = h^{-1}(\lang(G_{\mathrm{ex}}))\).
It is easy to see that every \(t \in L\) is of the form
\begin{equation*}
  \label{eq:sigmaspine}
  \sigma\bigl(\#, u_1\#, x\bigr) \csub \sigma\bigl(v_1\#, u_2\#, x\bigr) \csub \cdots \csub \sigma\bigl(v_{n-1}\#, u_n\#, x\bigr) \csub \sigma\bigl(v_{n}\#, \#, \#\bigr)
\end{equation*}
for some \(n \geq 1\), and \(u_i \in (ca^*c)^+\), \(v_i \in (db^*d)^+\), for each \(i \in [n]\).
In general, given a tree \(t\) of the form
\begin{equation}
  \label{eq:sigmaspine2}
  \sigma\bigl(v_1\#, u_1\#, x\bigr) \csub \cdots \csub \sigma\bigl(v_n\#, u_n\#, \zeta\bigr) \qquad \text{with } n \geq 1\,\text{,} \quad \zeta \in \{\#\} \cup X\,\text{,}
\end{equation}
where \(v_i \in (db^*d)^+\) and \(u_i \in (ca^*c)^+\), \(i \in [n]\), we will call the monadic subtrees \(u_{j}\) (resp.\ \(v_j\)) of \(t\) the \emph{\(a\)-chains} (resp.\ the \emph{\(b\)-chains}) of \(t\).  A \emph{chain} is either an \(a\)- or a \(b\)-chain.  The rightmost root-to-leaf path in \(t\) (that is labeled \(\sigma \cdots \sigma \zeta\)) will be referred to as \(t\)'s \emph{spine.}

For every tree \(t\) of the form as in \eqref{eq:sigmaspine2}, we let \(\iota(t) = v_1 u_1^R v_2 u_2^R \cdots v_n u_n^R\).
We view \(\Gamma\) as a parenthesis alphabet, such that \(b\) acts as right inverse to \(a\), and \(d\) to \(c\).  Then \(\iota(t)\) is a Dyck word, for every \(t \in L\).
\begin{proposition}
  \label{prop:iota-dyck}
  For every \(t \in L\), \(\iota(t) \in D_\Gamma^*\).
\end{proposition}
\begin{proof}
  Define \(\iota'\) for sentential forms of \(G_{\mathrm{ex}}\) by setting
  \[\iota'(\delta_1(v\#, \eta)) = v \iota'(\eta)\,\text{,} \quad \iota'(\delta_2(u \#, \eta)) = u^R \iota'(\eta)\,\text{,} \quad \iota(A(v\#, u\#, \eta)) = v u^R \iota'(\eta)\,\text{,}\]
  and \(\iota'(\#)=\epsilon\), for each \(\eta \in \TT{N_{\mathrm{ex}} \cup \Delta}{1}{0}\), and \(u\), \(v \in \Gamma^*\).  One can show by induction for every \(\eta \in \TT{N_{\mathrm{ex}} \cup \Delta}{1}{0}\) that if \(\eta_{\mathrm{ex}} \Rightarrow^*_{G_{\mathrm{ex}}} \eta\), then \(\iota'(\eta) \in D_\Gamma^*\).
  Moreover, \(\iota'(h(t)) \in D_\Gamma^*\) implies that \(\iota(t) \in D_\Gamma^*\), for any \(t \in \TT{\Delta}{1}{0}\).  This proves the proposition.
\end{proof}

There is the following relation between the numbers of symbol occurrences in \(t \in L\).

\begin{proposition}
  \label{prop:counts}
  For every \(t \in L\), 
  \(\abs{t}_c = \abs{t}_d = 4 \cdot \abs{t}_{\sigma} - 6\).
\end{proposition}
\begin{proof}
  One can show by induction for every \(\eta \in \TT{N_{\mathrm{ex}} \cup \Delta}{1}{0}\) that if \(\eta_{\mathrm{ex}} \Rightarrow^*_{G_{\mathrm{ex}}} \eta\), then \(\eta\) fulfills the equation
  \[\abs{\eta}_c = \abs{\eta}_d = 2 \cdot \abs{\eta}_{\delta_1} + 2 \cdot \abs{\eta}_{\delta_2} + 3 \cdot \abs{\eta}_A - 6 \,\text{.}\]
  Obviously, this property transfers to \(t \in L\) in the manner described above.
\end{proof}

Each chain of \(t \in L\) is uniquely determined by the other chains of \(t\), because \(\iota(t)\) is a Dyck word, and every chain contains either only symbols from \(\{a,c\}\), or only symbols from~\(\{b,d\}\).
\begin{observation}
  \label{obs:determinism}
  Let \(t \in L\), let \(w \in \pos(t)\) with \(t(w) \in \Gamma \cup \{\#\}\), and let \(s = t [x_1]_w\).  There is exactly one \(u \in \TT{\Gamma \cup \{\#\}}{1}{0}\) such that \(s \cdot u \in \lang(G_{\mathrm{ex}})\).
\end{observation}

\begin{example}
  The preimage of the tree from Example~\ref{ex:derivation} under \(h\) is
  \[t \; = \; \sigma(\#, cac^2a^2c \#, x) \csub \sigma(dbd\#, ca^2c\#, x) \csub \sigma(db^2d^2b^2d\#, \#,\#)\,\text{.}\]
  Obiously, \(\iota(t) = ca^2c^2acdbdca^2cdb^2d^2b^2d\), and it takes only a little patience to verify that \(\iota(t) \in D_\Gamma^*\).
\end{example}

In the following sections, we will prove that there is no cftg \(G\) with \(\lang(G) = L\).  Therefore, the following theorem holds.
\begin{theorem}
  \label{thm:main-theorem}
  The class of linear context-free tree languages is not closed under inverse linear tree homomorphisms.
\end{theorem}

\section{\boldmath A normal form for \(G\)}
\label{sec:normal-form}
Assume there is a cftg \(G = (N, \Sigma, \eta_0, P)\) such that \(\lang(G) = L\).  
In this section, we show (in a sequence of intermediate normal forms) that if \(G\) exists, then it can be chosen to be of a very specific form:
Let
\[t = \sigma(u_1\#, v_1\#, x) \csub \cdots \csub \sigma(u_n\#, v_n\#, \#) \in L\,\text{.}\]

If we consider the subtrees \(\sigma(u_i, v_i)\) as \emph{symbols} from an infinite alphabet \(\Lambda\), then
\(t\) can be understood as a word, and \(L\) as a word language, over \(\Lambda\).  In
fact, in the course of the next lemmas, we will see that therefore \(G\) can be
assumed to be of a form that is quite close to a context-free word grammar.
For example, in Lemma~\ref{lem:spine-grammar-five} it will be shown that the productions of G may be
assumed to be of the forms
\begin{enumerate}[\itshape (i)]
\item \(A \to B \cdot u\), with \(u \in \sTT{\Gamma}{p}{p}\),
\item \(A \to B \csub C\), and
\item \(A \to \sigma(x_i, x_j, x_{p+1})\),
\end{enumerate}
which correspond to \emph{(i)}~chain productions \(A \to B\), \emph{(ii)}~rank~2 productions \(A \to BC\) and
\emph{(iii)}~terminal productions \(A \to \sigma\) of context-free grammars.
In the next lemma, we start out with distinguishing nonterminals by whether they contribute to the spine of a tree or to its chains.

\begin{lemma}
  \label{lem:spine-grammar-two}
  We may assume for \(G\) that there is \(p \in \Nat\) such that \(N = N_s \cup N_c\) for two disjoint sets \(N_s = N_s^{(2p)}\) and \(N_c = N_c^{(p)}\).
  Moreover, \(\eta_0 = S(\#, \ldots, \#)\) for some \(S \in N_s\),
  and every production in \(P\) is of one of the following forms:\footnote{Recall that \(A(x_1, \ldots, x_{k})\) and \(A^{(k)}\) were identified!}
  \begin{multicols}{2}
    \begin{itemize}
    \item[\textit{(A1)}]
      \(A \to B\bigl(C_1, \ldots, C_p, D_1, \ldots, D_p\bigr)\),
    \item[\textit{(A2)}] \(A \to x_{p+q}\),
    \item[\textit{(A3)}] \(A \to \sigma(x_{i}, x_{j}, x_{p+q})\),
    \item[\textit{(A4)}]
      \(E \to F\bigl(C_1, \ldots, C_p\bigr)\),
    \item[\textit{(A5)}] \(E \to x_q\),
    \item[\textit{(A6)}] \(E \to \gamma(x_{q})\),
    \end{itemize}
  \end{multicols}
  \noindent where \(A\), \(B\), \(D_1\), \ldots, \(D_p \in N_s\), \(E\), \(F\), \(C_1\), \ldots, \(C_p \in N_c\), \(i\), \(j\), \(q \in [p]\), and \(\gamma \in \Gamma\).
\end{lemma}
\begin{proof}
  We begin by assuming that there is a number \(p \in \Nat\) such that \(N = N^{(p)}\), the productions in \(P\) are of the forms
  \begin{itemize}
  \item[\textit{(N1)}] \(A(x_1, \ldots, x_{p}) \to B(C_1(x_1, \ldots, x_p), \ldots, C_p(x_1, \ldots, x_p))\),
  \item[\textit{(N2)}] \(A(x_1, \ldots, x_{p}) \to x_i\) for some \(i \in [p]\),
  \item[\textit{(N3)}] \(A(x_1, \ldots, x_{p}) \to \gamma(x_i)\) for some \(\gamma \in \Gamma\) and \(i \in [p]\), or
  \item[\textit{(N4)}] \(A(x_1, \ldots, x_{p}) \to \sigma(x_i, x_j, x_q)\) for some \(i\), \(j\), \(q \in [p]\),
  \end{itemize}
  and that \(\eta_0 = S(\#, \ldots, \#)\) for some \(S \in N^{(p)}\).  This assumption comes without loss of generality: we may demand that \(G\) is in normal form \cite[Thm.~14]{Maibaum1974} and then introduce dummy parameters to make every nonterminal of rank \(p\).  One fixed parameter \(x_q\) can be used to store \(\#\) through the course of every derivation, then it is possible to use the production \(A \to x_q\) instead of \(A \to \#\).

  Let the regular tree grammar \(H = (Q, \Sigma, s, R)\) be given by \(Q = \{s, c\}\), and \(R\) contains the productions
  \begin{equation*}
    s \to \sigma(c, c, s) \; + \; \# \qquad \text{and} \qquad
    c \to \gamma(c) \; + \; \#
  \end{equation*}
  for every \(\gamma \in \Gamma\).

  We use Rounds's well-known method \cite{Rounds1970,Rounds1970a} to construct a cftg \(G' = (N', \Sigma, \eta_0', P')\) such that \(\lang(G') = \lang(G) \cap \lang(H)\).  Since \(\lang(G) \subseteq \lang(H)\), it is clear that \(\lang(G') = \lang(G)\).  However, as a side-effect of the method, \(G'\) is of the desired form.
  We describe the method briefly, in our own notation.  Let
  \(N' = \bigl\{  A_s^{(2p)} \mid A \in N^{(p)} \bigr\} \cup \bigl\{ A_c^{(p)} \mid A \in N^{(p)} \bigr\}\).
  
  Define two functions \(\Phi_s \colon \TT{N}{1}{p} \to \TT{N'}{1}{2p}\) and \(\Phi_c \colon \TT{N}{1}{p} \to \TT{N'}{1}{p}\) by simultaneous induction, such that
  \[\Phi_c(x_i) = x_i\,\text{,}\quad \text{and} \quad \Phi_s(x_i) = x_{p+i}\]
  for every \(x_i \in X_p\),
  and
  \begin{align*}
    \Phi_c(A(\eta_1, \ldots, \eta_p)) &= A_c \bigl(\Phi_c(\eta_1), \ldots, \Phi_c(\eta_p)\bigr)\,\text{,} \\
    \Phi_s(A(\eta_1, \ldots, \eta_p)) &= A_s \bigl(\Phi_c(\eta_1), \ldots, \Phi_c(\eta_p), \Phi_s(\eta_1), \ldots, \Phi_s(\eta_p)\bigr)
  \end{align*}
  for every \(A \in N\), and \(\eta_1\), \ldots, \(\eta_p \in \T_{N}(X_p)\).

  For every production \(A(x_1, \ldots, x_p) \to \eta\) in \(P\) of form \textit{(N1)} or \textit{(N2)}, the set \(P'\) contains the productions
  \( A_c(x_1, \ldots, x_{p}) \to \Phi_c(\eta)\) and \( A_s(x_1, \ldots, x_{2p}) \to \Phi_s(\eta)\).
  Moreover, for every production in \(P\) of form \textit{(N3)} (resp.\ \textit{(N4)}), \(P'\) contains the production
  \( A_c (x_1, \ldots, x_{p}) \to \gamma(\Phi_{c}(x_i)) = \gamma(x_i)\) (resp.\ \(A_s (x_1, \ldots, x_{2p}) \to \sigma(\Phi_{c}(x_i), \Phi_c(x_j), \Phi_s(x_q)) = \sigma(x_i, x_j, x_{p+q})\)).
  Let \(N_s = \{ A_s  \mid A \in N\}\) and \(N_c = \{ A_c  \mid A \in N\}\), and let \(\eta_0' =  S_s (\#, \ldots, \#)\).  Then it is easy to see that \(G'\) is of the form as demanded above.
\end{proof}

In the next step we show that we require at most two spine-producing nonterminals on the right-hand side of a production of \(G\).  The construction works by guessing beforehand which of the nonterminals of \(N_s\) in a production's right-hand side will eventually be chosen to contribute to the tree's spine.  

\begin{lemma}
  \label{lem:spine-grammar-three}
  We may assume for \(G\) that there is \(p \in \Nat\) such that \(N = N_c \cup N_s\) with \(N_c = N^{(p)}\) and \(N_s = N^{(p+1)}\).
  Moreover, \(\eta_0 = S(\#, \ldots, \#)\) for some \(S \in N_s\),  and every production of \(G\) is of one of the following forms:
  \begin{multicols}{2}
    \begin{itemize}
    \item[\textit{(B1)}] \(A \to B(C_1, \ldots, C_p, x_{p+1})\),
    \item[\textit{(B2)}] \(A \to B(x_1, \ldots, x_p, D)\),
    \item[\textit{(B3)}] \(A \to x_{p+1}\),
    \item[\textit{(B4)}] \(A \to \sigma(x_i, x_j, x_{p+1})\),
    \item[\textit{(B5)}] \(E \to F(C_1, \ldots, C_p)\),
    \item[\textit{(B6)}] \(E \to x_i\),
    \item[\textit{(B7)}] \(E \to \gamma(x_i)\),
    \item[~]
    \end{itemize}
  \end{multicols}
  \noindent where \(A\), \(B\), \(D \in N_s\), \(E\), \(F\), \(C_1\), \ldots, \(C_p \in N_c\), \(i\), \(j \in [p]\), and \(\gamma \in \Gamma\).
\end{lemma}
\begin{proof}
  Assume that \(G = (N, \Sigma, \eta_0, P)\) is of the form as given in Lemma~\ref{lem:spine-grammar-two}.  We will construct an equivalent cftg \(G''\) of the form demanded above.

  However, we construct first an intermediate cftg
  \(G' = (N', \Sigma, \eta_0', P')\),
  where \(N' = N_c \cup N_s' \cup \{S'\}\), such that \(S' \notin N_c \cup N_s'\) is a new nonterminal symbol, and
  \[N_s' = \bigl\{ \langle A,q \rangle^{(p+1)} \mid A \in N_s,\, q \in [p]\bigr\}\,\text{.} \]
  Moreover, \(\eta_0'=  S' (\#, \ldots, \#)\),
  and \(P'\) contains the productions
  \begin{enumerate}[\itshape (i)]
  \item \(\langle A, q\rangle \to \langle B, \tilde q \rangle\bigl(C_1, \ldots, C_p, \langle D_{\tilde q}, q \rangle\bigr)\)\\
    for every production of form \emph{(A1)}, and every \(q\), \(\tilde q \in [p]\);
  \item \(\langle A, q \rangle \to x_{p+1}\) for every production of form \emph{(A2)};
  \item \(\langle A, q \rangle \to \sigma(x_i, x_j, x_{p+1})\) for every production of form \emph{(A3)};
  \item every production of form \textit{(A4)}, \textit{(A5)}, or \textit{(A6)},
  \item \(S' \to \langle S, q \rangle\) for every \(q \in [p]\).
  \end{enumerate}

  We now prove that \(\lang(G') = \lang(G)\).  To this end, it is necessary to consider only OI derivations, as otherwise counting derivation steps becomes bothersome.
  It is easy to prove by induction for every \(n \in \Nat\), chain-producing nonterminal \(C \in N_c\) and \(t \in \TT{\Sigma}{1}{p}\) that \(C \derivOI{G}{n} t\) if and only if \(C \derivOI{G'}{n} t\).  

  Next, we show for every \(n \in \Nat\), \(q \in [p]\), \(A \in N_s\), and \(t \in \TT{\Sigma}{1}{p+1}\), that
  \[A \derivOI{G}{n} t \csub x_{p+q} \qquad \text{if and only if} \qquad \langle A, q \rangle \derivOI{G'}{n} t \csub x_{p+1} \,\text{.}\]
  The proof uses the fact that for every \(A \in N_s\) and \(t \in \lang(G, A)\), there is precisely one occurrence of a variable from \(\{x_{p+1}, \ldots, x_{2p}\}\) in \(t\).
  We proceed by complete induction on \(n\) (using Lemma~\ref{lem:technical} to decompose OI derivations).  The base case \(n = 0\) holds vacuously.  Continue by a case analysis on the production applied first in the derivation.  Let \(n \in \Nat\), \(A \in N_s\), \(q \in [p]\), and \(t \in \TT{\Sigma}{1}{p+1}\).  Assume that the production \(A \to B(C_1, \ldots, C_p, D_1, \ldots, D_p)\) is in \(P\).  Then
  \begin{alignat*}{2}
    &&& A \derivOI{G}{} B(C_1, \ldots, C_p, D_1, \ldots, D_p) \derivOI{G}{n} t \csub x_{p+q} \\
    &\text{iff}\quad&& \exists m \in \Nat\,\text{, } \tilde u \in \lTT{\Sigma}{1}{m}\,\text{, } \theta \in \Theta^{m}_{2p}\,\text{, } v \in \TT{\Sigma}{m}{2p} \colon \\
    &&&
    B \derivOI{G}{n_1} \tilde u \cdot \theta \,\text{,}\quad
    \theta \cdot [C_1, \ldots, C_p, D_1,\ldots, D_p] \derivOI{G}{n_2} v\,\text{,}\quad
    t \csub x_{p+q} = \tilde u \cdot v\,\text{,} \quad n = n_1 + n_2 \displaybreak[1] \\
    &\text{iff}&& \exists m \in \Nat\,\text{, } \tilde u \in \lTT{\Sigma}{1}{m+1}\,\text{, } \theta \in \Theta^{m}_{p}\,\text{, } v \in \TT{\Sigma}{m}{p}\,\text{, } \tilde q \in [p]\,\text{, } w \in \TT{\Sigma}{1}{p+1} \colon \tag{\(\dag\)} \\
    &&&
    B \derivOI{G}{n_1} \tilde u \cdot [\theta, x_{p+\tilde q}] \,\text{,}\quad
    \theta \cdot [C_1, \ldots, C_p] \derivOI{G}{n_2} v\,\text{,}\quad
    D_{\tilde q}  \derivOI{G}{n_3} w \csub x_{p+q}\,\text{,} \\
    &&&
    t = u \cdot [v, w]\,\text{,}\quad n = n_1 + n_2 +n_3 \displaybreak[1] \\
    &\text{iff}&& \exists m \in \Nat\,\text{, } \tilde u \in \lTT{\Sigma}{1}{m+1}\,\text{, } \theta \in \Theta^{m}_{p}\,\text{, } v \in \TT{\Sigma}{m}{p}\,\text{, } \tilde q \in [p]\,\text{, } w \in \TT{\Sigma}{1}{p+1} \colon \\ 
    &&&
    \langle B, \tilde q \rangle \derivOI{G'}{n_1} \tilde u \cdot [\theta, x_{p+1}] \,\text{,}\quad
    \theta \cdot [C_1, \ldots, C_p] \derivOI{G'}{n_2} v\,\text{,}\quad
    \langle D_{\tilde q}, q \rangle  \derivOI{G'}{n_3} w \csub x_{p+1} \,\text{,} \\
    &&&
    t = u \cdot [v, w]\,\text{,}\quad
    n = n_1 + n_2 + n_3 \displaybreak[1] \\
    &\text{iff}&& \langle A, q\rangle \derivOI{G'}{} \langle B, \tilde q \rangle (C_1, \ldots, C_p, \langle D_{\tilde q}, q\rangle) \derivOI{G'}{n} t\,\text{.}
  \end{alignat*}
  To understand why direction ``only if'' holds at point (\(\dag\)) above, observe that at this point, \(\pi_m \cdot v\) has the form \(w \csub x_{p+q}\), for some \(w \in \TT{\Sigma}{1}{p+1}\).  Since \(\pi_m \cdot v\) is generated by \(\pi_{\theta(m)} \cdot [C_1, \ldots, C_p, D_1, \ldots, D_p]\), there is some \(\tilde q \in [p]\) such that \(D_{\tilde q} \derivOI{G}{*} w \csub x_{p+q}\).
  
  If the production \(A \to x_{p+q}\) is in \(P\), with \(q \in [p]\), then
  \(A \derivOI{G}{} x_{p+1} \csub x_{p+q}\) if and only if \(\langle A, q\rangle \derivOI{G'}{} x_{p+1}\)
  by construction.  
  Finally, if \(A \to \sigma(x_i,x_j,x_{p+q})\) is in \(P\), then \(A \derivOI{G}{} \sigma(x_i,x_j, x_{p+1}) \csub x_{p+q}\) if and only if \(\langle A, q\rangle \derivOI{G'}{} \sigma(x_i,x_j, x_{p+1})\).
  \sepstars
  So for every \(t \in \TT{\Sigma}{1}{2p}\), we have that \(t \in \lang(G,S)\) if and only if there is some \(q \in [p]\) such that \(t \csub x_{p+1} \in \lang(G',\langle S,q \rangle)\).
  
  Let \(s \in \TSigma\).
  Then \(s \in \lang(G)\) if and only if there is \(t \in \lang(G,S)\) such that \(s = t \cdot \langle \#, \ldots, \# \rangle\),  and by the above, this is equivalent to \(t \csub x_{p+1} \in \lang(G',\langle S,q \rangle)\) for some \(q \in [p]\).  By use of the productions \textit{(v)}, this holds precisely if \(t \csub x_{p+1} \in \lang(G',S')\), i.e.\ \(s \in \lang(G')\).  Therefore, \(\lang(G) = \lang(G')\).

  The cftg \(G''\) results from \(G'\) by replacing every production of form \(A \to B(C_1, \ldots,C_p, D)\) in \(P'\) by the two productions
  \[A \to B_{C_1\cdots C_p}(x_1, \ldots, x_p, D) \quad \text{and} \quad B_{C_1\cdots C_p} \to B(C_1, \ldots, C_p, x_{p+1})\]
  for some new nonterminal \(B_{C_1\cdots C_p}\) of \(G''\).  It is easy to see that \(\lang(G'') = \lang(G')\), so a formal proof is omitted.
\end{proof}

The next normal form shows that the form of a chain of \(t \in L\) is already determined on the spine of \(t\).  We can therefore omit chain-producing nonterminals.

\begin{lemma}
  \label{lem:normal-form-1}
  We may assume that \(G\) is of the form \(G = (N, \Sigma, \eta_0, P)\), such that \(N = N^{(p+1)}\) for some \(p \in \Nat\), \(\eta_0 = S(\#, \ldots, \#)\) for some \(S \in N\)
  and the productions in \(P\) are of the forms\\[12pt]
  \vspace{12pt}
  \begin{minipage}[t]{1.0\linewidth} 
    \begin{multicols}{2}
      \begin{itemize}
      \item[\itshape (C1)] \(A \to B \cdot u\),\;
        where \(u \in \sTT{\Gamma}{p}{p}\),
      \item[\itshape (C2)] \(A \to B \csub C\),
      \item[\itshape (C3)] \(A \to x_{p+1}\),
      \item[\itshape (C4)] \(A \to \sigma(x_i,x_j, x_{p+1})\),\;
        where \(i\), \(j \in [p]\),
      \end{itemize}
    \end{multicols}
  \end{minipage}
  \noindent and where \(A\), \(B\), \(C \in N\).
\end{lemma}
\begin{proof}
  Assume that \(G\) is of the form given in Lemma~\ref{lem:spine-grammar-three}. 
  Moreover, we may assume that \(\lang(G,E) \ne \emptyset\), by Lemma~\ref{lem:cftg-total}.  The construction preserves our normal form.

  Note that for every \(E \in N_c\), we have \(\lang(G,E) \subseteq \TT{\Gamma}{1}{p}\).  Choose some fixed tree \(u_E \in \lang(G, E)\) for each \(E \in N_C\), and 
  let \(n\), \(m \in \Nat\).
  Given \(\eta \in \TT{N \cup \{\#\}}{n}{m}\), we define \(\phi(\eta) \in \TT{N' \cup \Sigma}{n}{m}\) as follows.  If \(n \ne 1\), let \(\phi(\eta) = \bigl\langle \phi(\pi_1 \cdot \eta), \ldots, \phi(\pi_n \cdot \eta) \bigr\rangle\).
  If \(n = 1\), let
  \begin{alignat*}{2}
    \phi(A \cdot \eta) &= A \cdot \phi(\eta) & \qquad\quad & \text{for every } A \in N_s \text{ and } \eta \in \TT{N}{p+1}{m}\,\text{,}\\
    \phi(E \cdot \eta) &= u_E \cdot \phi(\eta) && \text{for every } E \in N_c \text{ and } \eta \in \TT{N}{p}{m}\,\text{,}\\
    \phi(x_q) &= x_q && \text{for every } q \in [m]\,\text{, and} \\
    \phi(\#) &= \#\,\text{.} &&
  \end{alignat*}

  We construct the cftg \(G' = (N', \Sigma, \eta_0, P')\) with \(N' = \{A^{(p+1)} \mid A \in N_s\}\), and \(P'\) contains the productions
  \begin{enumerate}[\itshape (i)]
  \item \(A \to B(\phi(C_1), \ldots, \phi(C_p), x_{p+1})\) for every production of form \textit{(B1)} in \(P\),
  \item \(A \to B \csub D\) for every production of form \textit{(B2)} in \(P\),
  \item \(A \to x_{p+1}\) for every production of form \textit{(B3)} in \(P\),
  \item \(A \to \sigma(x_i, x_j, x_{p+1})\) for every production of form \textit{(B4)} in \(P\).
  \end{enumerate}
  Observe that in \textit{(i)}, \(\phi(C_i) \in \TT{\Sigma}{1}{p}\) for each \(i \in [p]\).

  \paragraph{(\(\pmb{\supseteq}\))} To prove that \(\lang(G') \subseteq \lang(G)\), we show for every \(n \in \Nat\), \(A \in N'\), and \(t \in \TT{\Sigma}{1}{p+1}\) that
  \[A \Rightarrow^n_{G'} t  \qquad \text{implies} \qquad A \Rightarrow^*_{G} t\,\text{.}\]
  The induction base holds trivially.  We continue with the following case analysis.  Let \(n \in \Nat\) and \(t \in \TT{\Sigma}{1}{p+1}\).
  
  \paragraph{(I)} Let
  \[A \Rightarrow_{G'} B(\phi(C_1), \ldots, \phi(C_p), x_{p+1}) \Rightarrow^n_{G'} t \cdot \langle \phi(C_1), \ldots, \phi(C_p), x_{p+1}\rangle\]
  for some production \(A \to B(C_1, \ldots, C_p, x_{p+1})\) in \(P\).
  By the induction hypothesis, \(B \Rightarrow^*_G t\), and clearly \(C_i \Rightarrow^*_G \phi(C_i)\) for each \(i \in [p]\), therefore
  \[
    A \Rightarrow_G B \cdot \langle C_1, \ldots, C_p, x_{p+1}\rangle
    \Rightarrow^*_G t \cdot \langle \phi(C_1), \ldots, \phi(C_p), x_{p+1}\rangle\,\text{.}
  \]
  
  \paragraph{(II)} Let
  \(A \Rightarrow_{G'} B \csub D \Rightarrow_{G'}^n t \)
  for some production \(A \to B \csub D\) in \(P\).
  Then there are \(n_1\), \(n_2 \in \Nat\), \(u\) and \(v \in \TT{\Sigma}{1}{p+1}\) such that
  \[B \Rightarrow^{n_1}_{G'} u\,\text{,} \quad D \Rightarrow^{n_2}_{G'} v\,\text{,} \quad n = n_1 + n_2\,\text{,} \quad \text{and} \quad t = u \csub v\,\text{.}\]
  By the induction hypothesis, we have that
  \(B \Rightarrow^*_{G} u\) and  \( D \Rightarrow^*_{G} v\),
  and therefore
  \[
    A \Rightarrow_G B \csub D \Rightarrow_G^* u \csub v \,\text{.}
  \]

  \paragraph{(III)} Let \(A \Rightarrow_{G'} x_{p+1}\).  This means that also \(A \Rightarrow_G x_{p+1}\).

  \paragraph{(IV)}  Let \(A \Rightarrow_{G'} \sigma(x_i, x_j, x_{p+1})\).  Then \(A \Rightarrow_G \sigma(x_i, x_j, x_{p+1})\).
  \sepstars
  Let \(s \in \lang(G')\).  Then there is some \(t \in \lang(G',S)\) such that \(s = t \cdot \langle \#, \ldots, \# \rangle\).  By the above, \(t \in \lang(G,S)\), and therefore \(s \in \lang(G)\).  Thus, \(\lang(G') \subseteq \lang(G)\).

  \paragraph{(\(\pmb{\subseteq}\))}  We continue the proof of correctness with the direction \(\lang(G) \subseteq \lang(G')\).  It rests on the following property.  For every \(n \in \Nat\), \(A \in N_s\), \(\eta \in \TT{N \cup \{\#\}}{p+1}{0}\), \(s \in \lTT{\Sigma}{1}{1}\), and \(t \in \TT{\Sigma}{1}{0}\), if
  \[\eta_0 \Rightarrow_G^* s \cdot A \cdot \eta \Rightarrow_G^n s \cdot t \,\text{,} \qquad \text{then also}
    \qquad
    A \cdot \phi(\eta) \Rightarrow^*_{G'} t\,\text{.}
  \]
  The induction base holds vacuously, so again we continue with a case analysis.  Let \(n \in \Nat\), \(s \in \lTT{\Sigma}{1}{1}\), and \(t \in \TT{\Sigma}{1}{0}\).

  \paragraph{(I)} Let
  \(\eta_0 \Rightarrow_G^* s \cdot A \cdot \eta \Rightarrow_G s \cdot B(C_1, \ldots, C_p, x_{p+1}) \cdot \eta \Rightarrow_G^n s \cdot t\).
  Then
  \[
    A \cdot \phi(\eta) \Rightarrow_{G'}
    B \cdot \langle \phi(C_1), \ldots, \phi(C_p), x_{p+1} \rangle \cdot \phi(\eta)
    = B \cdot \phi(\langle C_1, \ldots, C_p, x_{p+1} \rangle \cdot \eta)\,\text{,}
  \]
  and by the induction hypothesis,
  \(B\cdot \phi(\langle C_1, \ldots, C_p, x_{p+1} \rangle \cdot \eta) \Rightarrow_{G'}^* t\).

  \paragraph{(II)} Let
  \(\eta_0 \Rightarrow_G^* s \cdot A \cdot \eta \Rightarrow_G s \cdot B(x_1, \ldots, x_p, D) \cdot \eta \Rightarrow_G^n s \cdot t\).
  Then
  \[
    A \cdot \phi(\eta) \Rightarrow_{G'} B(x_1, \ldots, x_{p}, D) \cdot \phi(\eta)
    = B \cdot \phi(\langle x_1, \ldots, x_{p}, D \rangle \cdot \eta) \,\text{,}
  \]
  and by the induction hypothesis,
  \(B \cdot \phi(\langle x_1, \ldots, x_{p}, D \rangle \cdot \eta) \Rightarrow_{G'}^* t\).

  \paragraph{(III)} Let
  \(\eta_0 \Rightarrow_G^* s \cdot A \cdot \eta \Rightarrow_G s \cdot \pi_{p+1} \cdot \eta \Rightarrow_G^n s \cdot t\)
  by the production \(A \to x_{p+1}\).  Then
  \[A \cdot \phi(\eta) \Rightarrow_{G'} x_{p+1} \cdot \phi(\eta) = \phi(\pi_{p+1} \cdot \eta)\,\text{.}\]
  If \(\pi_{p+1} \cdot \eta = \#\), then 
  \(\phi(\pi_{p+1} \cdot \eta) = \# = t\).
  Otherwise, \(\pi_{p+1} \cdot \eta = B \cdot \kappa\) for some \(B \in N_s\) and \(\kappa \in \TT{N \cup \{\#\}}{p+1}{0}\).
  By the induction hypothesis,
  \(\phi(B \cdot \kappa) = B \cdot \phi(\kappa) \Rightarrow_{G'}^* t\).

  \paragraph{(IV)} Let \(u\), \(v \in \Gamma^*\) such that
  \[\eta_0 \Rightarrow_G^* s \cdot A \cdot \eta \Rightarrow_G s \cdot \sigma(\pi_i \cdot \eta, \pi_j \cdot \eta, \pi_{p+1} \cdot \eta) \Rightarrow_G^n s \cdot \sigma(u \#, v \#, t)\,\text{.}\]
  As in case (III), either \(\pi_{p+1} \cdot \eta = \#\), and then \(\phi(\pi_{p+1} \cdot \eta) = t\), or otherwise \(\pi_{p+1} \cdot \eta = B \cdot \kappa\) with \(B \cdot \phi(\kappa) \Rightarrow^*_{G'} t\).

  Moreover, as \(s \cdot \sigma(u\#, v\#, t) \in \lang(G)\), Observation~\ref{obs:determinism} entails that \(\lang(G, \pi_i \cdot \eta) = \{u\#\}\) and \(\lang(G, \pi_j \cdot \eta) = \{v\#\}\), from which we conclude that \(\phi(\pi_i \cdot \eta) = u \#\) and \(\phi(\pi_j \cdot \eta) = v \#\). So
  \[A \cdot \phi(\eta) \Rightarrow_{G'} \sigma\bigl(u \#, v \#, \phi(\pi_{p+1} \cdot \eta)\bigr) \Rightarrow^*_{G'} \sigma(u \#, v \#, t)\,\text{.}\]
  \vspace{-2\belowdisplayskip} \sepstars
  Let \(t \in \lang(G)\).  Then \(\eta_0 = S \cdot \langle \#, \ldots, \# \rangle \Rightarrow_G^* t\).
  The above property entails that
  \[S \cdot \phi(\langle \#, \ldots, \# \rangle) = S \cdot \langle \#, \ldots, \# \rangle \Rightarrow_{G'}^* t\,\text{,}\] and hence \(t \in \lang(G')\).
\end{proof}

It turns out, to derive the spine of \(t \in L\), no projecting productions \(A \to x_i\) are required: since \(G\) is close to a context-free word grammar with productions \textit{(C1)}~\(A \to B\), \textit{(C2)}~\(A \to BC\), \textit{(C3)}~\(A \to \epsilon\) and \textit{(C4)}~\(A \to \sigma\), we can eliminate the productions of form \textit{(C3)} by using the well-known method to remove \(\epsilon\)-productions from context-free grammars.

\begin{lemma}
  \label{lem:spine-grammar-five}
  In Lemma~\ref{lem:normal-form-1}, it is no restriction to demand that \(G\) has no productions of the form~\textit{(C3)}.
\end{lemma}
\begin{proof}
  Let \(Q = \{A \in N \mid A(x_1, \ldots, x_{p+1}) \Rightarrow_G^* x_{p+1}\}\)  
  and construct the cftg \(G' = (N, \Sigma, \eta_0, P')\), where \(P'\) contains all productions from \(P\) of forms \textit{(C1)}, \textit{(C2)} and \textit{(C4)}.
  Moreover, for every production of form \textit{(C2)}, \(P'\) contains the productions
  \[A \to B \quad\text{if}\quad C \in Q\,\text{,} \qquad \text{ and } \qquad
    A \to C \quad\text{if}\quad B \in Q\,\text{.}\]
  Observe that both productions are of form \textit{(C1)}.

  We prove that \(\lang(G') = \lang(G)\).  For the direction \(\lang(G') \subseteq \lang(G)\), we show for every \(n \in \Nat\), \(A \in N\), and \(t \in \TT{\Sigma}{1}{p+1}\), that if \(A \Rightarrow^n_{G'} t\), then also \(A \Rightarrow^*_{G} t\).  The proof is by complete induction on \(n\).  The induction base is trivial; we proceed by a case analysis on the form of the production applied first.
  
  Assume that 
  \(A \Rightarrow_{G'} B \Rightarrow^{n}_{G'} t\).
  From the induction hypothesis, \(B \Rightarrow^*_G t\).  There are three subcases.  Either, the production \(A \to B\) is in \(P\), in which case \(A \Rightarrow_G^* t\).  Otherwise, by construction, there is some production \(A \to B \csub C\) or \(A \to C \csub B\) in \(P\) such that \(C \Rightarrow^*_G x_{p+1}\).  If it is \(A \to B \csub C\) (the other case is analogous), we have
  \[A \Rightarrow_G B \csub C \Rightarrow_G^* B \Rightarrow_G^* t \,\text{.}\]
  
  The property is trivially true if the applied production is from \(P\).  For every \(s \in \lang(G')\), there is some \(t \in \lang(G',S)\) with \(s = t \cdot \langle \#, \ldots, \# \rangle\).  By the above, \(S \Rightarrow_{G}^* t\), and therefore \(s \in \lang(G)\).
  \sepstars
  It remains to show the direction \(\lang(G) \subseteq \lang(G')\).  We show for every \(n \in \Nat\), \(A \in N\), and \(t \in \TSigma(X_{p+1})\), that if \(A \Rightarrow^n_{G} t\) and \(t \ne x_{p+1}\), then also \(A \Rightarrow^*_{G'} t\).  The induction base is trivial, so again we continue by a case analysis on derivations of nonzero length.  Let \(n \in \Nat\), \(A \in N\), and \(t \in \TSigma(X_{p+1})\) with \(t \ne x_{p+1}\).

  If \(A \Rightarrow_{G} B \csub C \Rightarrow^n_G t\), then there are \(n_1\), \(n_2 \in \Nat\) and \(t_1\), \(t_2 \in \TT{\Sigma}{1}{p+1}\) with \(t = t_1 \csub t_2\),
  \(B \Rightarrow^{n_1}_G t_1\), \( C \Rightarrow^{n_2}_G t_2\), and \(n = n_1 + n_2\).
  If neither \(t_1 = x_{p+1}\) nor \(t_2 = x_{p+1}\), then by the induction hypothesis, also
  \[A \Rightarrow_{G'} B \csub C \Rightarrow_{G'}^* t_1 \csub t_2 = t\,\text{.}\]
  If precisely one of \(t_1\) and \(t_2\) is equal to \(x_{p+1}\) (say \(t_1 = x_{p+1}\), the other case is analogous), then the production \(A \to C\) is in \(P'\).  So, with the induction hypothesis,
  \[A \Rightarrow_{G'} C \Rightarrow_{G'}^* t_2 = t\,\text{.}\]
  The case \(t_1 = t_2 = x_{p+1}\) is precluded by the assumption that \(t \ne x_{p+1}\).

  Similarly, the case that the first production is of form \textit{(C3)} is precluded by the assumption on \(t\).  For any other production, the proof goes through without surprises.

  Now let \(s \in \lang(G)\).  Then there is \(t \in \lang(G, S)\) such that \(s = t \cdot \langle \#, \ldots, \# \rangle\).  Note that \(t \ne x_{p+1}\), because \(\# \notin \lang(G)\).  Thus by the above property, \(S \Rightarrow^*_{G'} t\), and therefore \(s \in \lang(G')\).
\end{proof}

Finally, it is convenient to remove the torsions from productions of the form \textit{(C1)}.  Then whenever \(A \Rightarrow_G^* B \cdot u\), we know that \(u\) is torsion-free.  The construction works by guessing which torsion will be applied in the next derivation step, and pre-arranging this torsion in the tuple of the current production.  However, there is a price to pay: we must now allow for torsions in ``branching'' productions \(A \to B \cdot \theta_1 \csub C \cdot \theta_2\).  

\begin{lemma}
  \label{lem:normal-form}
  We may assume that \(G\) is of the form \(G = (N, \Sigma, \eta_0, P)\), such that \(N = N^{(p+1)}\) for some \(p \in \Nat\), \(\eta_0 = S(\#, \ldots, \#)\) for some \(S \in N\)
  and the productions in \(P\) are of the forms 
  \begin{itemize}
  \item[\itshape (D1)] \(A \to B \cdot u\),\;
    where \(u \in \lsTT{\Gamma}{p}{p}\),
  \item[\itshape (D2)] \(A \to B \cdot \theta_1 \csub C \cdot \theta_2\),\;
    where \(\theta_1\), \(\theta_2 \in \sTheta{p}{p}\),
  \item[\itshape (D3)] \(A \to \sigma(x_i,x_j, x_{p+1})\),\;
    where \(i\), \(j \in [p]\),
  \end{itemize}
  and where \(A\), \(B\), \(C \in N\).
\end{lemma}
\begin{proof}
  Assume that \(G\) is as in Lemma~\ref{lem:spine-grammar-five}.
  Construct a new cftg \(G' = (N', \Sigma, \eta_0', P')\), where \(N' = N \times \sTheta{p}{p} \cup \{S'\}\) for some distinct nonterminal \(S'\), \(\eta_0' = S'(\#, \ldots, \#)\), and \(P'\) contains the productions
  \begin{enumerate}[\itshape (i)]
  \item \(A^{\theta'} \to B^{\theta}\cdot s\) 
    for every production of form \textit{(C1)} and \(\theta \in \sTheta{p}{p}\), where \(\lin(\theta \cdot u) = (s, \theta')\);
  \item \(A^{\Id{p+1}} \to B^{\theta_1} \cdot \theta_1 \csub C^{\theta_2} \cdot \theta_2\)
    for every production of form \textit{(C2)}, and \(\theta_1\), \(\theta_2 \in \sTheta{p}{p}\);
  \item \(A^{\Id{p+1}} \to \sigma(x_i, x_j, x_{p+1})\)
    for every production of form \textit{(C4)};
  \item  \(S' \to S^{\theta}\)
    for every \(\theta \in \sTheta{p}{p}\).
  \end{enumerate}

  To prove the construction correct, we demonstrate for every \(n \in \Nat\), \(A \in N\), \(v \in \sTT{\Gamma}{p}{p}\), and \(t \in \TT{\Sigma}{1}{p+1}\), that
  \[A \cdot v \Rightarrow_{G}^n t \qquad \text{if and only if} \qquad \exists \theta \in \sTheta{p}{p} \colon \; A^{\theta} \cdot \theta \cdot v \Rightarrow_{G'}^n t\,\text{.}\]
  The proof is by complete induction on \(n\).  The induction base holds trivially, hence we proceed by a case analysis on derivations of nonzero length.  Assume therefore that \(n \in \Nat\), \(A \in N\), \(v \in \sTT{\Gamma}{p}{p}\), and \(t \in \TT{\Sigma}{1}{p+1}\).

  \paragraph{(I)} By construction, \(A \cdot v \Rightarrow_G \sigma(\pi_i \cdot v, \pi_j \cdot v, x_{p+1})\) if and only if \(A^{\Id{p+1}} \cdot v \Rightarrow_{G'} \sigma(\pi_i \cdot v, \pi_j \cdot v, x_{p+1})\).

  \paragraph{(II)}  Assume that \(A \cdot v \Rightarrow_G B \cdot v \csub C \cdot v \Rightarrow^{n}_G t\).
  Then there are \(n_1\), \(n_2 \in \Nat\), \(t_1\), and \(t_2 \in \TT{\Sigma}{p+1}{p+1}\) such that
  \[B \cdot v \Rightarrow_G^{n_1} t_1\,\text{,} \quad C \cdot v \Rightarrow_G^{n_2} t_2\,\text{,} \quad t = t_1 \csub t_2\,\text{,} \quad \text{and} \quad n = n_1 + n_2\,\text{.}\]
  By induction, there are \(\theta_1\), \(\theta_2 \in \sTheta{p}{p}\) such that \(B^{\theta_1} \cdot \theta_1 \cdot v \Rightarrow_{G'}^{n_1} t_1\) and \(C^{\theta_2}\cdot \theta_2 \cdot v \Rightarrow_{G'}^{n_2} t_2\).  Thus,
  \[A^{\Id{p+1}} \cdot v \Rightarrow_{G'} B^{\theta_1} \cdot \theta_1 \cdot v \csub C^{\theta_2} \cdot \theta_2 \cdot v \Rightarrow_{G'}^{n_1 + n_2} t_1 \csub t_2 = t\,\text{.}\]
  Conversely, let \(\theta_1\), \(\theta_2\), \(\theta_3 \in \sTheta{p}{p}\), and \(n_1\), \(n_2 \in \Nat\) such that
  \[
    A^{\theta_1} \cdot \theta_1 \cdot v \Rightarrow_{G'} B^{\theta_2} \cdot \theta_2 \cdot \theta_1 \cdot v \csub C^{\theta_3} \cdot \theta_3 \cdot \theta_1 \cdot v \Rightarrow_{G'}^n t\,\text{.}
  \]
  By construction, \(\theta_1 = \Id{p+1}\).  Moreover, there are \(n_1\), \(n_2 \in \Nat\), \(t_1\), and \(t_2 \in \TT{\Sigma}{p+1}{p+1}\) such that
  \[B^{\theta_2} \cdot \theta_2 \cdot v \Rightarrow_{G'}^{n_1} t_1\,\text{,} \quad C^{\theta_3} \cdot \theta_3 \cdot v \Rightarrow_{G'}^{n_2} t_2\,\text{,} \quad t = t_1 \csub t_2\,\text{,} \quad \text{and} \quad n = n_1 + n_2\,\text{.}\]
  By the induction hypothesis, \(B \cdot v \Rightarrow_G^{n_1} t_1\) and \(C \cdot v \Rightarrow_G^{n_2} t_2\), thus
  \[A \cdot v \Rightarrow_{G} B \cdot v \csub C \cdot v \Rightarrow_{G}^{n_1 + n_2} t_1 \csub t_2 = t\,\text{.}\]

  \paragraph{(III)} Assume finally that \(A \cdot v \Rightarrow_G B \cdot u \cdot v \Rightarrow_G^n t\) for some \(n \in \Nat\). By the induction hypothesis, there is some \(\theta \in \sTheta{p}{p}\) such that \(B^{\theta} \cdot \theta \cdot u \cdot v \Rightarrow_{G'}^n t\).  By construction, there is a production \(A^{\theta'} \to B^{\theta} \cdot s\), where \(s \in \lsTT{\Gamma}{p}{p}\) with \(s \cdot \theta' = \theta \cdot u\).  Thus we have
  \[A^{\theta'} \cdot \theta' \cdot v \Rightarrow_{G'} B^{\theta} \cdot s \cdot \theta' \cdot v = B^{\theta}\cdot \theta \cdot u \cdot v \Rightarrow_{G'}^n t\,\text{.}\]

  For the other direction, let \(A^{\theta'} \cdot \theta' \cdot v \Rightarrow_{G'} B^{\theta} \cdot s \cdot \theta' \cdot v \Rightarrow_{G'}^n t\) for some \(n \in \Nat\), \(\theta\) and \(\theta' \in \sTheta{p}{p}\).  By construction, there is some production \(A \to B \cdot u\) in \(P\), such that \(s \cdot \theta' = \theta \cdot u\).  Hence, \(B^{\theta} \cdot s \cdot \theta' \cdot v = B^{\theta}\cdot \theta \cdot u \cdot v \Rightarrow_{G'}^n t\).  By the induction hypothesis, \(B \cdot u \cdot v \Rightarrow_G^n t\).
  Thus also \(A \cdot v \Rightarrow_G B \cdot u \cdot v \Rightarrow_G^{n} t\).
  \sepstars
  Let \(t \in \TT{\Sigma}{1}{0}\).  Then \(t \in \lang(G)\) if and only if \(t \in \lang(G,S \cdot  \langle \#, \ldots, \# \rangle)\).  By the above property, this holds precisely if there is some \(\theta \in \sTheta{k}{k}\) such that \(t \in \lang(G, S^\theta \cdot \theta\cdot \langle \#, \ldots, \# \rangle)\), and it is easy to see that \(S^\theta \cdot \theta \cdot \langle \#, \ldots, \# \rangle = S^\theta \cdot \langle \#, \ldots, \# \rangle\).  By construction of \(G'\), we receive that
  \[t \in \lang(G) \quad \text{iff} \quad t \in \lang(G,S \cdot  \langle \#, \ldots, \# \rangle) \quad \text{iff} \quad t \in \lang(G',S' \cdot  \langle \#, \ldots, \# \rangle) \quad \text{iff} \quad t \in \lang(G')\,\text{,}\]
  and therefore \(\lang(G) = \lang(G')\).
\end{proof}

\begin{quote}
  \em  Assume for the rest of this work that there is a cftg \(G\) of the form stated in Lemma~\ref{lem:normal-form} such that \(\lang(G) = L\).  Let \(\chi\) denote the tuple \(\langle \#, \ldots, \# \rangle\).  Then \(\eta_0 = S \cdot \chi\).
\end{quote}

\section{Derivation trees}
\label{sec:derivation-trees}
A derivation of a tree \(t \in \lang(G)\) can be described faithfully by a binary tree \(\kappa\).\footnote{As a subset \(\pos(\kappa) \subseteq \{1,2\}^*\) that is prefix-closed and such that \(w 1 \in \pos(\kappa)\) iff \(w 2 \in \pos(\kappa)\).}  These \emph{derivation trees} will help us analyze the structure of the derivations in \(G\).

Formally, let \(\kappa\) be a binary tree such that each position \(\delta \in \pos(\kappa)\) is equipped with
two nonterminal symbols \(A_\delta\) and \(B_\delta \in N\),
a torsion-free tuple \(s_\delta \in \lsTT{\Gamma}{p}{p}\),
a torsion \(\theta_\delta \in \sTheta{p}{p}\),
and two numbers \(i_\delta\) and \(j_\delta \in [p]\).
Then \(\kappa\) is an \emph{\((A_\epsilon,\theta_\epsilon)\)-derivation tree} if for every \(\delta \in \pos(\kappa)\),
\begin{enumerate}[\itshape (i)]
  \setlength{\itemsep}{0pt}
\item \(A_\delta \Rightarrow_G^* B_\delta \cdot s_\delta\),
\item if \(\delta\) is a leaf of \(\kappa\), then the production \(B_\delta \to \sigma(x_{i_\delta}, x_{j_\delta}, x_{p+1})\) is in \(P\),
\item otherwise, \(B_\delta \to A_{\delta 1} \cdot \theta_{\delta 1} \csub A_{\delta 2} \cdot \theta_{\delta 2}\) is a production in \(P\).
\end{enumerate}
Let \(t \in \TT{\Sigma}{1}{p+1}\).  We say that \(\kappa\) is an \emph{\((A_\epsilon,\theta_\epsilon)\)-derivation tree of \(t\)} (or: \(\kappa\) \emph{derives} \(t\)) if either
\(\kappa\) has only one node and \(t = \sigma(x_{i_\epsilon}, x_{j_\epsilon}, x_{p+1}) \cdot s_\epsilon \cdot \theta_\epsilon\), or, otherwise,
there are \(t_1\), \(t_2 \in \TT{\Sigma}{1}{p+1}\) such that \(\kappa \vert_1\) derives \(t_1\), \(\kappa \vert_2\) derives \(t_2\), and \(t = (t_1 \csub t_2) \cdot s_\epsilon \cdot \theta_\epsilon\).
An \((S, \Id{p+1})\)-derivation tree (of \(t\)) will simply be called a \emph{derivation tree (of \(t\)).}

There is the following relation between derivations and derivation trees.
\begin{proposition}
  \label{prop:derivation-trees}
  Let \(t \in \TT{\Sigma}{1}{p+1}\), let \(A \in N\), and \(\theta \in \sTheta{p}{p}\).  Then \(A \cdot \theta \Rightarrow_G^* t\) if and only if there is an \((A, \theta)\)-derivation tree of \(t\).
\end{proposition}
\begin{proof}
  Straightforward by complete induction on \(\abs{t}_\sigma\).
\end{proof}
As a direct corollary, \(t \cdot \chi \in L\) if and only if there is a derivation tree of \(t\).
We close our discussion of derivation trees with the following pumping lemma. 
It states that if there is some \(s_\delta\) in \(\kappa\) which has a sufficiently large component, then an iterable pair of nonterminals occurs in the derivation of \(s_\delta\).

\begin{quote}
  \em In the sequel, fix the pumping number \(H = \card{N} \cdot h_{\mathrm{max}}\), where \(h_{\mathrm{max}}\) is the maximal size of a component of \(u\) in a production of \(G\) of form \textit{(D1)}.
\end{quote}
\begin{lemma}
  \label{lem:pumping}
  Let \(\kappa\) be a derivation tree and \(\delta \in \pos(\kappa)\).  If there are \(i \in [p]\) and \(w\), \(w' \in \Gamma^*\) such that \(\pi_i \cdot s_\delta = w' w x_{i}\) and \(\abs{w} > H\), then there exist \(v\), \(y\), \(z \in \lsTT{\Sigma}{p}{p}\) such that
  \begin{multicols}{2}
    \begin{itemize}
    \item[\itshape (i)] \(s_\delta = v \cdot y \cdot z\),
    \item[\itshape (ii)] \(\pi_i \cdot y \cdot z\)
      is a suffix of \(w x_i\),
    \item[\itshape (iii)] \(\abs{\pi_i \cdot y} > 0\), and
    \item[\itshape (iv)] for each \(j \in \Nat\),
      \(A_\delta \Rightarrow_G^* B_\delta \cdot v \cdot
      y^j \cdot z\).
    \end{itemize}
  \end{multicols}
\end{lemma}
\begin{proof}
  By definition of \(\kappa\), \(A_\delta \Rightarrow_G^* B_\delta \cdot s_\delta\).  So there are \(n \in \Nat\), \(C_1\), \ldots, \(C_n \in N\) and \(e^{(1)}\),~\ldots,~\(e^{(n)} \in \lsTT{\Gamma}{p}{p}\) such that
  \[C_1 \cdot e^{(1)} \Rightarrow_G C_2 \cdot e^{(2)} \cdot e^{(1)} \Rightarrow_G \cdots \Rightarrow_G C_n \cdot e^{(n)} \cdots e^{(1)} \]
  where \(C_1 = A_\delta\), \(C_n = B_\delta\), \(e^{(1)} = \Id{p+1}\), and \(e^{(n)} \cdots e^{(1)} = s_\delta\).
  
  If \(C_1\), \ldots, \(C_n\) are pairwise distinct, then \(n \leq \card{N}\) and the maximal size of a component of \(s_\delta\) is \(\card{N} \cdot h_{\mathrm{max}} = H\), which contradicts the assumption that \(\abs{w} > H\).
  We can therefore choose two indices \(\ell\), \(k \in [q]\) with \(\ell < k\) such that \(C_\ell = C_k\), the size of \(\pi_i \cdot e^{(k)} \cdots e^{(\ell+1)}\) is nonzero, and \(\ell\) and \(k\) are the two smallest numbers with these properties.
  Let
  \[v = e^{(n)} \cdots e^{(k+1)}\,\text{,} \quad y = e^{(k)} \cdots e^{(\ell+1)}\,\text{,} \quad \text{and} \quad z = e^{(\ell)} \cdots e^{(1)}\,\text{.} \]
  Then for every \(j \in \Nat\),
  \[A_\delta \cdot \Id{p+1} \Rightarrow_G^* C_\ell \cdot z \Rightarrow_G^* C_k \cdot y^j \cdot z \Rightarrow_G^* B_\delta \cdot v \cdot y^j \cdot z\,\text{.}\]
  Moreover, the size of \(\pi_i \cdot y \cdot z\) is at most \(H\), therefore \(\pi_i \cdot y \cdot z\) is a suffix of \(w x_i\).
\end{proof}

\section{Dyck words and sequences of chains}
\label{sec:dyck-words-sequences}
This section prepares some necessary notions for the upcoming counterexample.  We introduce a sequence \(U_1\), \(U_2\), \ldots{} of Dyck words.  Later, an element of this sequence will contribute to the chains of the tree \(t\) used in the counterexample.  As described in the introduction, the proof revolves around the factorization of \(t\) into trees \(t_1\) and \(t_2\) that is induced by the derivation of \(t\).  So we will analyze the corresponding factorizations of the Dyck words \(U_i\).

Moreover, we will introduce here the notion of \emph{defects}, which can be understood as the ``unclosed parentheses'' in \(t_1\), resp.\ \(t_2\).  Finally, a lemma on \emph{perturbations} is given, which will be used to show that if the defects in \(t_1\) are modified (or: perturbed), then the word formed by the chains of the resulting tree lies in another Dyck congruence class.  This implies that the resulting tree does not ``fit together'' with \(t_2\) any longer.

First of all, let us fix the following constants.    
Let \(q = 2p\), and let \(m = 2^{q-1} + 1\).
For every \(i \in \Nat\), let \(\alpha_i = c a^{i m H} c\) and \(\beta_i = d b^{i m H} d\).
Define the sequence \(U_1\), \(U_2\), \ldots{} of words over \(\Gamma\) by
\[U_1 = \alpha_1 \beta_1 \qquad \text{and} \qquad U_{i+1} = \alpha_{i+1} U_i U_i \beta_{i+1} \quad \text{for every } i \geq 1\,\text{.}\]
We make the following observation.
\begin{observation}
  \label{obs:form-of-U}
  For every \(i \geq 1\),
  \begin{enumerate}[\itshape (i)]
  \item \(U_i \in D_\Gamma^*\), and
  \item \(U_i = u_1 v_1 \cdots u_{n} v_{n}\), where \(n = 2^{i-1}\), \(u_j \in (ca^+c)^+\), and \(v_j \in (db^+d)^+\), for \(j \in [n]\).
  \end{enumerate}
\end{observation}
For each \(U_i\) of the above form, let \(Z_i = \langle u_1^R, v_1, \ldots, u_n^R, v_n \rangle\).  The components \(u_\ell^R\) and \(v_\ell\) of \(Z_i\) will also be called \emph{chains}, as later on they will end up as the chains of some \(t \in L\).
For every factorization of \(Z_i\) into \[Z_i' = \langle u_1^R, v_1, u_2^R, v_2, \ldots, u_j^R\rangle \quad \text{and} \quad Z_i'' = \langle v_j, u_{j+1}^R, v_{j+1}, \ldots, u_n^R, v_n \rangle\,\text{,} \quad j \in [n]\,\text{,}\]
consider the respective factors \(P_{i,j} = u_1 v_1 u_2 v_2 \cdots u_j\) and \(S_{i,j} = v_j u_{j+1} v_{j+1} \cdots u_n v_n\) of \(U_i\).
\begin{proposition}
  \label{prop:decomposition}
  The factors \(P_{i,j}\) and \(S_{i,j}\) can be written as
  \begin{equation}
    \label{eq:decomposition}
    P_{i,j} = \alpha_i V_{i-1} \alpha_{i-1} \cdots V_1 \alpha_1 \qquad \text{and} \qquad S_{i,j} = \beta_1 W_1 \cdots \beta_{i-1} W_{i-1} \beta_i\,\text{,}
  \end{equation}
  such that \(V_\ell\), \(W_\ell \in \{\epsilon, U_\ell\}\) and \(V_\ell \ne W_\ell\) for every \(\ell \in [i-1]\).
\end{proposition}
\begin{proof}
  By induction on \(i\).  The base case \(U_1 = \alpha_1 \beta_1\) has only one factorization, \(P_{1,1} = \alpha_1\) and \(S_{1,1} = \beta_1\), which fulfills the property.  Let \(i \geq 1\) and consider \(U_{i+1} = \alpha_{i+1} U_i U_i \beta_{i+1}\).  A factorization \(P_{i+1,j} S_{i+1,j}\) of \(U_{i+1}\) induces a factorization of either the first or the second occurrence of \(U_i\) into, say \(P_{i,j'}\) and \(S_{i,j'}\) for some \(j' \in [2^{i-1}]\).  Therefore, \(U_{i+1} = \alpha_{i+1} V_i P_{i,j'} S_{i,j'} W_i \beta_{i+1}\) for \(V_i\), \(W_i \in \{\epsilon, U_i\}\) with \(V_i \ne W_i\).  By induction, \(P_{i,j'} = \alpha_i V_{i-1} \alpha_{i-1} \cdots V_1 \alpha_1\), and therefore \(P_{i+1,j} = \alpha_{i+1} V_i \alpha_i V_{i-1} \alpha_{i-1} \cdots V_1 \alpha_1\), for \(V_i\), \ldots, \(V_1\) as given above.  The same kind of argument works for~\(S_{i+1,j}\).
\end{proof}

Assume a factorization of \(U_i\) into \(P_{i,j}\) and \(S_{i,j}\) as given in~\eqref{eq:decomposition}. Then we denote by \(D_{i,j}\) the word
\[\$ \alpha_i V_{i-1}' \alpha_{i-1} \cdots V_1' \alpha_1 \$ \beta_1 W_1' \cdots \beta_{i-1} W_{i-1}' \beta_i \$\]
over \(\Gamma \cup \{\$\}\), where for every \(\ell \in [i-1]\),
\(V_\ell' = \$\) if \(V_\ell = U_\ell\), and \(V_\ell' = \epsilon\) if \(V_\ell = \epsilon\), and, analogously, \(W_\ell' = \$\) if \(W_\ell = U_\ell\), and \(W_\ell' = \epsilon\) if \(W_\ell = \epsilon\).
Let \(\ell\), \(k \in \Nat\) with \(\ell \leq k\).  We say that a word \(\gamma = \alpha_\ell \cdots \alpha_k\) (resp.\ \(\gamma = \beta_\ell \cdots \beta_k\)) is an \emph{\(a\)-defect} (resp.\ a \emph{\(b\)-defect}) \emph{in \(D_{i,j}\)} if \(\$ \gamma^R \$\) (resp.\ \(\$\gamma\$\)) occurs in \(D_{i,j}\).  When the factorization is clear, the reference to \(D_{i,j}\) is omitted.
Both \(a\)-defects and \(b\)-defects will be called \emph{defects.}
A chain in \(Z_i\) whose suffix is a defect is called a \emph{critical chain.}
\begin{proposition}
  \label{prop:defects}
  Consider a factorization of \(U_i\) into \(P_{i,j}\) and \(S_{i,j}\).
  \begin{enumerate}[\itshape (1)]
  \item There is no \(\ell \in [i]\) such that \(\alpha_\ell\) (or \(\beta_\ell\)) occurs in two distinct defects.
  \item The number of defects in \(D_{i,j}\) is \(i+1\).
  \item Each \(a\)-defect (resp.\ \(b\)-defect) is the suffix of some chain \(u_n\) (resp.\ \(v_n\)) in \(Z_i\), with \(n \in [2^{i-1}]\).
  \end{enumerate}
\end{proposition}
\begin{proof}
  For \textit{(1)}, observe that the \(a\)-defects in \(D_{i,j}\) are disjoint (non-overlapping) factors of the word \(\alpha_1 \cdots \alpha_i\).  A similar observation can be made for the \(b\)-defects in \(D_{i,j}\).
  For \textit{(2)}, it is easy to see from Proposition~\ref{prop:decomposition} that there are exactly \(i+2\) occurrences of the symbol \(\$\) in \(D_{i,j}\).  So there are \(i+1\) factors of the form \(\$ \gamma \$\) in \(D_{i,j}\), for \(\gamma \in \Gamma^*\).  By \textit{(1)}, the defects are pairwise distinct, so \(D_{i,j}\) contains precisely \(i+1\) defects.

  Regarding \textit{(3)}, let \(\gamma = \alpha_\ell \cdots \alpha_k\), \(k \geq \ell\), be an \(a\)-defect in \(D_{i,j}\) and let
  \[D_{i,j} = D' \$ \underbrace{\alpha_k \cdots \alpha_\ell}_{\gamma^R} \$ D'' \qquad \text{for some }  D', D'' \in (\Gamma \cup \{\$\})^*\,\text{.}\]
  By definition of \(D_{i,j}\), \(P_{i,j}\) is of the form
  \[P_{i,j} = P' U_k \alpha_k \cdots U_\ell \alpha_\ell P'' \qquad  \text{for some } P', P'' \in \Gamma^*\]
  if \(k < i\), and \(P_{i,j} = \alpha_k \cdots U_\ell \alpha_\ell P''\) if \(k = i\).
  As \(U_k\) ends with \(\beta_k\), \(\gamma\) is the suffix of some chain \(u_n\) in \(Z_i\).  A similar argument can be made if \(\gamma\) is a \(b\)-defect.
\end{proof}

Let \(P\), \(P' \in (ca^*c)^*\).  We say that \(P'\) is a \emph{perturbation} of \(P\) if it results from \(P\) by modifying the exponents of \(a\) in \(P\).
More precisely, let \(P\) be of the form
\[P = w_0 a^{f_1} w_1 \cdots w_{\ell-1} a^{f_\ell} w_\ell\,\text{,}\]
such that \(\ell \in \Nat\), \(w_0\), \ldots, \(w_\ell \in c^*\), and for each \(i \in [\ell]\), \(f_i > 0\).
Then \(P' \in \Gamma^*\) is called a \emph{perturbation} of \(P\) if
\[P' = w_0 a^{f_1'} w_1 \cdots w_{\ell-1} a^{f_\ell'} w_\ell\,\text{,}\]
for some \(f_1'\), \ldots, \(f_\ell' \in \Nat\).  The only perturbation of \(\epsilon\) is \(\epsilon\) itself.
\begin{lemma}
  \label{lem:key-lemma}
  Consider a factorization of \(U_i\) into \(P_{i,j}\) and \(S_{i,j}\), and let \(P_{i,j}'\) be a perturbation of \(P_{i,j}\), i.e.\
  \begin{equation}
    P_{i,j} = \alpha_i V_{i-1} \alpha_{i-1} \cdots V_1 \alpha_1 \qquad \text{and} \qquad P_{i,j}' = \alpha_i' V_{i-1}' \alpha_{i-1}' \cdots V_1' \alpha_1'\,\text{.}\label{eq:key}
  \end{equation}
  Then \(P'_{i,j} \equiv P_{i,j}\) if and only if \(V_{\ell}' \equiv \epsilon\) for every \(\ell \in [i-1]\) and \(\alpha_\ell' = \alpha_\ell\) for every \(\ell \in [i]\).
\end{lemma}
\begin{proof}
  The direction ``if'' is trivial.  For the other direction, we first prove for every \(i > 0\) and every perturbation \(U_i'\) of \(U_i\) that either \(U_i' \equiv \epsilon\) or \(U_i' = cXd\) for some \(X \not\equiv \epsilon\).  The proof is by induction on \(i\).  For the base case, consider a perturbation \(U_1' = ca^pcdb^qd\) of \(U_1\), where \(p\), \(q \in \Nat\).  Since \(U_1' \equiv ca^pb^qd\), \(U_1' \not\equiv \epsilon\) implies that \(p \ne q\), and therefore \(a^pb^q \not\equiv \epsilon\).  Assume now a perturbation
  \[P_{i+1}' = c a^p c P_i' P_i'' d b^q d\,\text{,} \qquad p,\,q \in \Nat\,\text{,}\]
  of \(P_{i+1}\), where \(P_i'\) and \(P_i''\) are perturbations of \(P_i\).  If \(P_{i+1'} \not \equiv \epsilon\), then either \(P_i'P_i''\equiv \epsilon\) and \(p \ne q\) as above.  Otherwise \(P_i'P_i''\) is of the form \(cXd\) with \(X \not \equiv \epsilon\).  But then \(P_{i+1}'\) is also of this form.

  We can now prove the direction  ``only if''.  Let
  \(P_{i,j}' \equiv P_{i,j}\).
  As \(V_\ell \in \{U_\ell, \epsilon\}\) for every \(\ell \in [i-1]\), \(P_{i,j}\) reduces to \(\alpha_i \cdots \alpha_1\).
  Assume that there is some \(\ell \in [i-1]\) with \(V_\ell' \not\equiv \epsilon\).  Then the reduction of \(P_{i,j}'\) would contain an occurrence of \(c\), by the property shown above.  But this is in contradiction to the assumption that \(P_{i,j}' \equiv P_{i,j}\).  Hence, \(V_1'\), \ldots, \(V_{i-1}' \equiv \epsilon\).  Then clearly also \(\alpha_\ell' = \alpha_\ell\) for every \(\ell \in [i]\).
\end{proof}
Let us remark that an analogous lemma can be formulated for perturbations of \(S_{i,j}\).  However, we will only consider perturbations of \(P_{i,j}\) afterwards.

\section{\boldmath A witness for \(\lang(G) \ne L\)}
\label{sec:witness-langg-ne}
In this section, we choose a tree \(t \in L\) whose chains form a sufficiently large word \(U_i\).  By viewing a derivation tree \(\kappa\) of \(t\), which induces a factorization \(t = t_1 \csub t_2\), we will see that the pumping lemma from Section~\ref{sec:derivation-trees} can be applied, and this leads to a perturbation in the defects of \(t_1\).  By Lemma~\ref{lem:key-lemma} right above, we receive the desired contradiction.

\begin{algorithm}[t]
  \caption{Derivation of \(h(t)\) in \(G_{\mathrm{ex}}\)}
  \label{alg:derivation}
  \algblockdefx[REPEAT]{REPEAT}{ENDREPEAT}
  [1]{\textbf{repeat} #1 \textbf{times}}
  {\textbf{end repeat}}
  \begin{algorithmic}
    \State\vphantom{\(A\)}\(\eta \gets \eta_{\mathrm{ex}}\)
    \ForAll{\(j \in \{1, \ldots, q\}\)}
      \REPEAT{\(j \cdot m \cdot H\)}
      \State \(\eta \gets \mathrm{apply}\bigl(A \to A(a x_1, b x_2, x_3), \eta  \bigr)\)
      \ENDREPEAT
      \State \(\eta \gets \mathrm{apply}\bigl(A \to A(ccx_1, d\#, A(c\#, ddx_2, x_3)), \eta  \bigr)\)
    \EndFor
    \State \(\eta \gets \mathrm{apply}\bigl(A \to \delta_2(cx_1, \delta_1(dx_2, x_3)), \eta  \bigr)\)
  \end{algorithmic}
\end{algorithm}
Let \(Z_q = \langle u_1, v_1, \ldots, u_{m-1}, v_{m-1} \rangle\), recalling that \(m = 2^{q-1} + 1\).  Moreover, let 
\[t = \sigma\bigl( \#, u_1\#, x \bigr) \csub \sigma\bigl(v_1 \#, u_2\#, x \bigr) \csub \cdots \csub \sigma\bigl(v_{m-2} \#, u_{m-1}\#, x \bigr) \csub \sigma\bigl( v_{m-1}\#, \#, \# \bigr)\,\text{.}\]
Observe that \(t\) contains \(m\) occurrences of \(\sigma\), and that \(\iota(t) = U_q\).
Moreover, the chains of \(t\) are of the form \(\alpha_1 \cdots \alpha_\ell\), resp.\ \(\beta_1 \cdots \beta_\ell\), for some \(\ell \in [q]\).

\begin{proposition}
 \(t \in L\).
\end{proposition}
\begin{proof}
  Algorithm~\ref{alg:derivation} demonstrates how to derive \(h(t)\) in \(G_{\mathrm{ex}}\).
  There, \(\mathrm{apply}(\pi, \eta)\) denotes the parallel application of the production \(\pi\) in \(P_{\mathrm{ex}}\) to every possible position in the sentential form \(\eta\).
  Clearly, the result is \(\eta = h(t)\), and therefore \(t \in L\).
\end{proof}

As \(m > 1\), there are \(t_1\), \(t_2 \in \TT{\Sigma}{1}{1}\) such that
\[
  A_\epsilon \cdot \chi \Rightarrow^*_G B_\epsilon \cdot s_\epsilon \cdot \chi
  \Rightarrow_G \bigl( A_1 \cdot \theta_1 \cdot s_\epsilon \csub A_2 \cdot \theta_2 \cdot s_\epsilon \bigr) \cdot \chi \Rightarrow_G^* t_1 \csub t_2 = t\,\text{.}
\]
Since both \(t_1\) and \(t_2\) contain at least one occurrence of \(\sigma\), there is a \(j \in [m-1]\) such that
\begin{align*}
t_1 &= \sigma(\#, u_1 \#, x) \csub \sigma(v_1\#, u_2 \#, x) \csub \cdots \csub \sigma(v_{j-1}\#, u_j \#, x) \quad \text{and} \\
t_2  &= \sigma(v_{j}\#, u_{j+1} \#, x) \csub \cdots \csub  \sigma(v_{m-2} \#, u_{m-1} \#, x) \csub \sigma(v_{m-1}\#, \#, \#)\,\text{,}
\end{align*}
and this factorization of \(t\) induces an according factorization of \(Z_q\) into \(Z'\) and \(Z''\) with
\[Z' =  \langle u_1, v_1, \ldots, u_j\rangle \qquad \text{and} \qquad Z'' = \langle v_j, \ldots, u_{m-1}, v_{m-1}\rangle \,\text{.}\]

\begin{example}
  Let us consider an example which relates the introduced concepts.
  Figure~\ref{fig:defect} displays the critical chain \(u_k\) in \(t_1\), whose defect is \(\gamma = \alpha_4 \alpha_5\).  As \(u_k\) is critical, every \(a\)-chain \(u_{k'}\) in \(t_1\) to its right (i.e., with \(k' > k\)) is of the form \(\alpha_1 \cdots \alpha_\ell\), for some \(\ell \leq 3\).  In our intuition, \(\gamma\) is a sequence of opening parentheses which have no corresponding closing parenthesis in~\(t_1\).  Therefore, \(t_2\) must contain a suitable sequence of closing parentheses.
  Formally, \(\gamma^R\) occurs in \(P_{i,j}\) as
  \[P_{i,j} = P' U_{5} \alpha_5 \alpha_4 U_3 P''\,\text{,} \qquad \text{so} \qquad D_{i,j} = D' \$ \gamma^R \$ D''\,\text{,}\]
  for some \(P'\), \(P'' \in \Gamma^*\) and \(D'\), \(D'' \in (\Gamma \cup \{\$\})^*\).  Therefore, \(\gamma\) is indeed a defect by definition. 
\end{example}

By Proposition~\ref{prop:defects}(2), the number of defects in \(D_{q,j}\) is \(q + 1 = 2p + 1\).  Thus either \(t_1\) contains at least \(p + 1\) critical chains, or \(t_2\) does.
\begin{quote}
  \em For the rest of this work, assume that \(t_1\) contains at least \(p+1\) critical chains.  The proofs for the other case are obtained mainly by substituting \(b\) for \(a\) and \(\beta\) for \(\alpha\).
\end{quote}

\begin{figure}
  \centering
  \begin{tikzpicture}[node distance=3ex and 3ex, text height=1.5ex, text depth=0.25ex, every node/.style={inner sep=1pt,minimum width=2.5ex,minimum height=2.5ex}]

  \begin{scope}[every on chain/.style={join}, start chain=1]
    \node [on chain=1] (s0){\(\sigma\)};
    \node [on chain=1] {\(\hphantom{M}\cdots\hphantom{M}\)};
    \node [on chain=1] (s1) {\(\sigma\)};
    \node [on chain=1] (cdots) {\(\hphantom{M}\hphantom{M}\cdots\hphantom{M}\hphantom{M}\)};
    \node [on chain=1] (s2) {\(\sigma\)};
    \node [on chain=1] {\(\hphantom{M}\cdots\hphantom{M}\)};
    \node [on chain=1] {\(x_1\)};
  \end{scope}

  \node[left=of s0] {\(t_1 =\)};

  \begin{scope}[start chain=2 going below,node distance=3pt]
    \node [on chain=2, below right=2pt and 2pt of s1] (c) {\(\alpha_1\)};
    \node [on chain=2] {\(\alpha_2\)};
    \node [on chain=2] (cU) {\(\alpha_3\)};
    \node [on chain=2] (g1) {\(\alpha_4\)};
    \node [on chain=2] (g2) {\(\alpha_5\)};
    \node [on chain=2] (hash) {\(\#\)};
  \end{scope}

  \node[below=-5pt of hash,text depth=3ex] {\(\underbrace{\phantom{M}}_{u_k}\)};
  
  \draw[] ($(s1.south east) + (-1pt,1pt)$) -- ($(c.north west) + (1pt,-1pt)$);

  \begin{scope}[start chain=3 going below,node distance=3pt]
    \node [on chain=3, below left=2pt and 2pt of s2] (d) {\(\beta_1\)};
    \node [on chain=3] {\(\beta_2\)};
    \node [on chain=3] (dU) {\(\beta_3\)};
    \node [on chain=3] {\(\#\)};
  \end{scope}

  \draw[] ($(s2.south west) + (1pt,1pt)$) -- ($(d.north east) + (-1pt,-1pt)$);

  \begin{scope}[start chain=4 going below,node distance=3pt]
    \node [on chain=4, below right=2pt and 2pt of s2] (cr) {\(\alpha_1\)};
    \node [on chain=4] {\(\alpha_2\)};
    \node [on chain=4] {\(\alpha_3\)};
    \node [on chain=4] {\(\#\)};
  \end{scope}

  \node [below left=8pt and 2pt of s1] (v1) {\(\vdots\)};

  \node [below left=6pt and 2pt of s0] (v0) {\#};
  \node [below right=8pt and 2pt of s0] (v00) {\(\vdots\)};


  \draw[] ($(s2.south east) + (-1pt,1pt)$) -- ($(cr.north west) + (1pt,-1pt)$);

  \draw[] ($(s1.south west) + (0pt,1pt)$) -- ($(v1.north) + (4pt,5pt)$);

  \draw[] ($(s0.south west) + (0pt,1pt)$) -- ($(v0.north) + (4pt,3pt)$);
  \draw[] ($(s0.south east) + (0pt,1pt)$) -- ($(v00.north) + (-4pt,5pt)$);

  \draw [decorate,decoration={brace,amplitude=5pt,mirror},xshift=-0.4pt] ($(g1.north west) + (0,-2pt)$) -- ($(g2.south west) + (0,2pt)$) node[black,midway,xshift=-10pt] {\(\gamma\)};

  \begin{pgfonlayer}{background}
    \filldraw [line width=4mm,line join=round,black!10] (c)  rectangle (dU);
  \end{pgfonlayer}

  \node[below=18pt of cdots, color=black!70] {\(U_3\)};
\end{tikzpicture}
  \caption{Occurrence of a defect \(\gamma\) in the critical chain \(u_k\) of \(t_1\)}
  \label{fig:defect}
\end{figure}

By Proposition~\ref{prop:derivation-trees}, there are a \(\hat t \in \TT{\Sigma}{1}{p+1}\) with \(t = \hat t \cdot \chi\), and a derivation tree \(\kappa\) of \(\hat t\).
Note that the height of \(\kappa\) is at most \(m\).
Therefore \(\abs{\delta} < m\) for every \(\delta \in \pos(\kappa)\).  If \(\delta = i_1 \cdots i_d\), then we denote the prefix \(i_{1} \cdots i_{d-\ell} \) of \(\delta\) by \(\delta_\ell\), for every \(\ell \in [0,d]\).  In particular, \(\delta_0 = \delta\) and \(\delta_d = \epsilon\).

Let \(s \in \sTT{\Gamma}{p}{p}\) and \(w \in \Gamma^*\).  If there is no possibility of confusion, we will briefly say that \(w\) is a component of \(s\) if \(s\) has a component of the form \(w x_i\), for some \(i \in [p]\).

\begin{proposition}
  \label{prop:chains}
  Let \(u_i\) be an \(a\)-chain of \(t_1\), with \(i \in [j]\).  There is a leaf \(\delta\) of \(\kappa\) such that
  \[u_i = w_{0} \cdots w_d \,\text{,} \]
  where \(d = \abs{\delta}\), and 
 \(w_\ell\) is a component of \(s_{\delta_\ell}\), for \(\ell \in [0,d]\).  Moreover, \(\delta_{d-1} = 1\).
\end{proposition}
\begin{proof}
  Every leaf node of \(\kappa\) contributes exactly one occurrence of \(\sigma\) to \(t\).  So the chain \(u_i\) is contributed to \(t\) by \(\kappa\)'s \(i\)-th leaf node \(\delta\), when enumerated from left to right.  Let \(d = \abs{\delta}\).  By tracing the path from \(\delta\) to the root of \(\kappa\), we see that
  \[u_i\# = \pi_{j_{\delta_0}} \cdot s_{\delta_0} \cdot \theta_{\delta_0} \cdots s_{\delta_{d-1}} \cdot \theta_{\delta_{d-1}} \cdot s_\epsilon \cdot \chi\,\text{.} \]
  Therefore \(u_i = w_{0} \cdots w_d \), where
  \(w_\ell\) is a component of \(s_{\delta_\ell}\), for each \(\ell \in [0,d]\).
\end{proof}

In particular, \(w_d\) is a component of \(s_\epsilon\).  The next lemma is a consequence of the fact that \(s_\epsilon\) has only \(p\) components apart from \(x_{p+1}\).

\begin{lemma}
  \label{lem:pumpable}
  There is an \(a\)-defect \(\gamma\) whose critical chain is of the form \(w' w\) for some \(w'\), \(w \in \Gamma^*\) such that \(w\) is a component of \(s_\epsilon\), and \(\abs{\gamma} > \abs{w} + m H\).
\end{lemma}
\begin{proof}
  Since \(t_1\) contains more than \(p\) critical chains, by Proposition~\ref{prop:chains} there must be two critical chains, say \(u \gamma \alpha_i\) and \(u' \gamma' \alpha_j\), where \(\gamma \alpha_i\) and \(\gamma' \alpha_j\) are distinct \(a\)-defects with \(i < j\), such that
  \[u \gamma \alpha_i = w' w \qquad \text{and} \qquad u' \gamma' \alpha_j = w'' w \qquad \text{for some } w',\, w'' \in \Gamma^*\,\text{,}\]
  and some  component \(w\) of \(s_\epsilon\).

  Observe that \(\alpha_i\) is not a suffix of \(w\), as otherwise \(\alpha_i\) would be a suffix of \(\alpha_j\).  Therefore \(\abs{w} < \abs{\alpha_i}\), and hence
  \[\abs{w} + mH < \abs{\alpha_i} + mH = \abs{\alpha_{i+1}} \leq \abs{\alpha_{j}} \leq \abs{\gamma' \alpha_j}\,\text{.}\]
  So the \(a\)-defect \(\gamma' \alpha_j\) satisfies the properties in the lemma.
\end{proof}

\begin{theorem}
  There is some \(t' \in \lang(G) \setminus L\).
\end{theorem}
\begin{proof}
  Let \(\gamma\) be the \(a\)-defect from Lemma~\ref{lem:pumpable}.  Assume that \(\gamma\)'s critical chain in \(t_1\) is \(u_k\), where \(k \in [j]\). Then \(u_k = w_0 \cdots w_{d}\), where \(w_\ell\) is a component of \(s_\ell\), for each \(\ell \in [0,d]\).  Moreover, \(\abs{\gamma} > \abs{w_d} + m H\).  Let \(f\) be the largest number such that \(w_f \cdots w_d\) has \(\gamma\) as suffix, then \(f \in [0, \ldots, d-1]\), and there are \(w\), \(w' \in \Gamma^*\) such that \(w_f = w' w\) and
  \(\gamma = w w_{f+1} \cdots w_d\).

  Since \(d < m\) and \(\abs{w w_{f+1} \cdots w_{d-1}} > m H\), there is a \(\tilde w \in \{w, w_{f+1}, \ldots, w_{d-1}\}\) such that \(\abs{\tilde w} > H\).  In other words, there is an \(\ell \in [f, d-1]\) such that \(A_{\delta_\ell} \Rightarrow^*_G B_{\delta_\ell} \cdot s_{\delta_\ell}\), and there is some \(i \in [p]\) such that either \textit{(i)}~\(\ell=f\) and \(\pi_i \cdot s_{\delta_\ell} =  w' \tilde w x_i\), or \textit{(ii)}~\(\ell \ne f\) and \(\pi_i \cdot s_{\delta_\ell} =  \tilde w x_i\).  In both cases Lemma~\ref{lem:pumping} can be applied, and we receive that \(s_{\delta_\ell} = v \cdot y \cdot z\), and by pumping zero times, also \(A_{\delta_\ell} \Rightarrow^*_G B_{\delta_\ell} \cdot v \cdot z\).  Therefore a derivation tree \(\kappa'\) can be constructed from \(\kappa\) by replacing the tuple \(s_{\delta_\ell}\) by \(v \cdot z\).  As \(\delta_\ell\)  begins with the symbol \(1\), this alteration does only concern \(t_1\), thus \(\kappa'\) derives a tree \(\hat t' \in \TT{\Sigma}{1}{p+1}\) such that \(\hat t' \cdot \chi = t_1' \csub t_2\), for some \(t_1' \in \TT{\Sigma}{1}{1}\).  Denote \(\hat t' \cdot \chi\) by \(t'\).

  Let us compare the \(k\)-th \(a\)-chain \(u_k'\) of \(t_1'\) to \(u_k\).
  Assume that the \(i\)-th components of \(v\), \(y\), and \(z\) are, respectively, \(v'x_i\), \(y'x_i\) and \(z'x_i\).
  Then in case~\textit{(i)}, there is a \(w'' \in \Gamma^*\) such that \(v' = w' w''\), as \(y' z'\) is a suffix of \(w\).
  Therefore,
  \[u_k = w_1 \cdots w' \underbrace{w'' y' z' w_{f+1} \cdots w_d}_\gamma \quad \text{and} \quad u_k' = w_1 \cdots w' w'' z' w_{f+1} \cdots w_d\,\text{.} \]
  In case~\textit{(ii)},
  \[u_k = w_1 \cdots \underbrace{w w_{f+1} \cdots w_{\ell-1} v' y' z' w_{\ell+1} \cdots w_d}_\gamma \quad \text{and} \quad u_k' = w_1 \cdots w_{\ell-1} v' z' w_{\ell+1} \cdots  w_d\,\text{.}\]
  It is easy to see that \(\abs{t'}_\sigma = \abs{t}_\sigma\), as the shape of \(\kappa\) was not modified.
  Thus Proposition~\ref{prop:counts} implies that if \(t' \in L\), then also  \(\abs{t'}_c = \abs{t}_c\) and \(\abs{t'}_d = \abs{t}_d\).  In particular, \(y' \in a^*\).  Therefore, both in case~\textit{(i)} and~\textit{(ii)}, \(P_{i,j}' = \iota(t_1')\) is a perturbation of \(P_{i,j}\).  Say that \(P_{i,j}\) and \(P_{i,j}'\) are of the form as in~\eqref{eq:key}.
  Since \(\abs{y'} > 0\) by Lemma~\ref{lem:pumping}, at least one \(a\) was removed from the occurrence of \(\gamma^R\) in \(P_{i,j}\).  Therefore, there is some \(e \in [q]\) such that \(\alpha_e \ne \alpha_e'\).  By Lemma~\ref{lem:key-lemma}, therefore \(P_{i,j}' \not \equiv P_{i,j}\), and hence \[\iota(t')  \equiv \iota(t_1') \; \iota(t_2) \not\equiv \iota(t_1) \; \iota(t_2) \equiv \epsilon\,\text{.}\]
  So \(\iota(t') \notin D_\Gamma^*\), and by Proposition~\ref{prop:iota-dyck}, \(t' \notin L\).
\end{proof}

Therefore, there is no cftg \(G\) with \(\lang(G) = h^{-1}(\lang(G_\mathrm{ex}))\), and we have proven Theorem~\ref{thm:main-theorem}.

\section{Linear monadic context-free tree languages and inverse homomorphisms}
\label{sec:lmcftg}

In this section, we close the paper with the positive result announced in the introduction.

\begin{theorem}
  \label{thm:lm-cftg}
  The class of linear monadic context-free tree languages is closed under inverse linear tree homomorphisms.
\end{theorem}
We will prove this theorem in the remainder of this section.  As the constructions are not very difficult, we use a style that is not so formal.
Let us start out with recalling a normal form for lm-cftg given in \cite{Fujiyoshi2006}.

Let $G=(N,\Delta,\eta_0,P)$ be an lm-cftg.\footnote{Where \(\Sigma\) and \(\Delta\) will denote arbitrary ranked alphabets in the following, unless stated otherwise.}  We say that $G$ is in \textit{Greibach normal form} if \(\eta_0 = S\) for some \(S \in N^{(0)}\) and each production in $P$ is of one of the following forms:
\begin{itemize}
\item[\itshape (G1)] $A \to \alpha$ for some $A\in N^{(0)}$, $\alpha \in \Delta^{(0)}$,
\item[\itshape (G2)] $A \to \delta(B_1,\ldots,B_{i-1},\eta,B_{i+1},\ldots,B_k)$ for some \(A \in N^{(0)}\), and \(\eta\in \TT{N}{1}{0}\), or
\item[\itshape (G3)] $A(x) \to \delta(B_1,\ldots,B_{i-1},\eta,B_{i+1},\ldots,B_k)$ for some $A \in N^{(1)}$ and $\eta\in \lTT{N}{1}{1}$,
\end{itemize}
and \(k \in \Nat\), \(i \in [k]\), \(\delta \in \Delta^{(k)}\), \(B_1\), \ldots, \(B_{i-1}\), \(B_{i+1}\), \ldots, \(B_k\in N^{(0)}\).  Note that every Greibach cftg is nondeleting.

\begin{lemma}[{\cite[Theorem 4.3]{Fujiyoshi2006}}]
 For every lm-cftg $G$ there is an equivalent lm-cftg $G'$ in Greibach normal form.
\end{lemma} 

The following decomposition theorem admits proving Theorem~\ref{thm:lm-cftg} in a modular manner.

\begin{lemma}[{\cite[Lemma~10]{Arnold1980}}]
  Let \(h \colon \TSigma(X) \to \TDelta(X)\) be a linear tree homomorphism.  There are a linear alphabetic tree homomorphism \(\phi\), as well as elementary ordered tree homomorphisms \(\psi_1\), \ldots, \(\psi_k\) for some \(k \in \Nat\) such that
  \(h = \psi_k \circ \cdots \circ \psi_1 \circ \phi\).
\end{lemma}

So in order to show that the linear monadic context-free tree languages are closed under inverse linear tree homomorphisms, it suffices to show closure under these two restricted types.

\begin{lemma}
  The class of linear monadic context-free tree languages is closed under inverse linear alphabetic tree homomorphisms.
\end{lemma}
\begin{proof}
  Assume an lm-cftg \(G = (N, \Delta, S, P)\) in Greibach normal form, and let \(h \colon \TSigma(X) \to \TDelta(X)\) be a linear alphabetic tree homomorphism.
  Let \(H = (M, \Sigma, Z, R)\) be a regular tree grammar such that \(\lang(R) = \TSigma\), and \(M\) is disjoint from \(N\).
  We use the same idea as in \cite[Theorem~4.1]{Arnold1978} to construct an lm-cftg \(G' = (N', \Sigma, S, P')\) with \(\lang(G') = h^{-1}(\lang(G))\).  Let \(N' = N \cup M \cup \{E^{(1)}\}\) such that \(E \notin N \cup M\), and let \(P'\) be given as follows.
  \begin{enumerate}[\itshape (i)]
  \item For every production of type \textit{(G1)} in \(P\), \(P'\) contains \(A \to E(\alpha)\).
  \item For every production of type \emph{(G2)} in \(P\),
    if \(h(\sigma) = \delta(x_{j_1}, \ldots, x_{j_k})\) for some \(n \in \Nat\), \(\sigma \in \Sigma^{(n)}\), and \(j_1\), \ldots, \(j_k \in [n]\), then \(P'\) contains
    \(A \to E(\sigma(u_1, \ldots, u_n))\), where for each \(\ell \in [n]\),
    \[u_\ell =
      \begin{dcases*}
        B_m & if \(\ell = j_m\) for some \(m \ne i\), \\
        \eta & if \(\ell = j_i\), \\
        Z & if \(\ell \notin \{j_1, \ldots, j_n\}\).
      \end{dcases*}
    \]
  \item The analogous applies to every production of type \textit{(G3)}.
  \item For every \(n \in \Nat\) and \(\sigma \in \Sigma^{(n)}\) such that \(h(\sigma) = x_{j}\) for some \(j \in [n]\), \(P'\) contains the production \(E(x) \to \sigma(u_1, \ldots, u_n)\), where for each \(\ell \in [n]\),
    \[u_\ell =
      \begin{dcases*}
        x & if \(\ell = j\),\\
        Z & otherwise.
      \end{dcases*}
    \]
  \item \(P'\) contains the productions \(E(x) \to x\) and \(E(x) \to E(E(x))\).
  \end{enumerate}
  The equivalence of \(G\) and \(G'\) is shown by defining a tree homomorphism \(\phi \colon \T_{N \cup \{E\}}(X) \to \T_N(X)\) such that \(\phi(E) = x\), and \(\phi(A) = A\) for every \(A \in N\).  Then it is easy to prove by induction on the length on the derivations that for every \(\eta \in \TT{N \cup \{E\}}{1}{0}\) and \(t \in \TT{\Sigma}{1}{0}\) we have \(\eta \Rightarrow_{G'}^* t\) if and only if \(\phi(\eta) \Rightarrow^*_G h(t)\).
\end{proof}

\begin{lemma}
  The class of linear monadic context-free tree languages is closed under inverse elementary ordered tree homomorphisms.
\end{lemma}
\begin{proof}
For this purpose, let \(\Omega\) be a ranked alphabet such that \(\Omega\) and \(\{\delta_1, \delta_2, \sigma\}\) are disjoint.  Let \(\Sigma = \Omega \cup \{\sigma^{(k)}\}\) and \(\Delta = \Omega \cup \{\delta_1^{(n-k+1)}, \delta_2^{(k)}\}\) for some \(n\), \(k \in \Nat\).
Let \(h \colon \TSigma(X) \to \TDelta(X)\) be the elementary ordered tree homomorphism with
\[h(\sigma(x_1, \ldots, x_n)) = \delta_1(x_1, \ldots x_{\ell-1}, \delta_2(x_{\ell}, \ldots, x_{\ell+k-1}), x_{\ell+k}, \ldots, x_{n})\]
for some \(\ell \in [n+1]\), and \(h\) is the identity on \(\Omega\).

Assume an lm-cftg \(G = (N, \Delta, S, P)\).  We will construct an lm-cftg \(G'\) such that \(\lang(G') = h^{-1}(\lang(G))\). We proceed in several steps.

First, we construct an lm-cftg $G_3$ with $\lang(G_3)=\lang(G)$ that fulfills the following property \emph{(P)}: in the right-hand side of each production of $G_3$ the terminals $\delta_1$ and $\delta_2$ either occur together, directly beneath each other, or none of these terminals occurs.  Formally, we demand for every production \(u \to v\) of \(G_3\) that \(v(\epsilon) \ne \delta_2\) and for every \(w \in \pos(v)\),
\[v(w) = \delta_1 \qquad \text{iff} \qquad w \ell \in \pos(v) \quad \text{and} \quad v(w \ell) = \delta_2\,\text{.} \]

We assume that $G$ is in Greibach normal form.  Moreover, we can assume without loss of generality that
\begin{itemize}
\item[--] \(\lang(G) \subseteq h(\TSigma)\), \footnote{If it is not, one can apply a similar method to the one in \cite[Theorem~2]{Nederhof2012} to construct a cftg \(G'\) with \(\lang(G') = \lang(G) \cap h(\TSigma)\).  Note that \(G'\) is again linear and monadic.}
\item[--] that \(G\) is total (by Lemma~\ref{lem:cftg-total}), and
\item[--] that \(G\) has no unreachable nonterminal symbols, i.e.\ for every \(A \in N\), there are \(\eta \in \lTT{N \cup \Delta}{1}{1}\) and \(\kappa \in \TT{N \cup \Delta}{}{}\) such that \(S \Rightarrow_G^* \eta \cdot A \cdot \kappa\) (cf.\ \cite[Proposition~14]{Arnold1980}).
\end{itemize}
Then, the following property holds for $G$.

\begin{observation}[{cf. \cite[Lemma 17]{Arnold1978}}]
  \label{obs:shape}
  For all $A\in N$ and $t\in \TT{\Delta\cup N}{1}{1}$ with $A \Rightarrow_{G}^{*} t$ we have that $t$ has no subtree of one of the following shapes:
\begin{itemize}
\item[--] $\gamma \cdot [u, \delta_2 \cdot v, w]$ for some $\gamma\in\Delta \setminus \{\delta_{1}\}$
\item[--] $\delta_1 \cdot [u, \gamma \cdot v, w]$ for some $\gamma\in\Delta \setminus \{\delta_{2}\}$ such that \(u \in \TT{\Delta \cup N}{\ell-1}{1}\), or
\item[--] $\delta_1 \cdot [u, \delta_2 \cdot v, w]$ with \(u \in \TT{\Delta \cup N}{m}{1}\) and \(m \ne \ell-1\),
\end{itemize}
and where \(u\), \(v\), \(w \in \TT{\Delta \cup N}{}{}\).
\end{observation}

Let in the following
\[\tilde N = \{A \in N \mid \exists u \in \TT{\Sigma}{1}{k} \colon A \Rightarrow_G^* \delta_2 \cdot u\}\,\text{.}\]
As \(G\) is total and \(\lang(G) \subseteq h(\TSigma)\), the following observation can be made.

\begin{observation}
  \label{obs:nonterminals}
  Let \(A \in \tilde N\).  For every \(t \in \lang(G,A)\), we have \(t(\epsilon) = \delta_2\).
\end{observation}

We now construct the lm-cftg $G_1=(N_1,\Sigma,S,P_1)$ with $N_1=N\cup \{C_\rho\mid\rho\in P\}$ and the following productions in $P_1$:
\begin{enumerate}[\itshape (i)]
 \item Every production $A\to t$ in $P$ with $t(\epsilon)\neq\delta_2$ is also in $P_1$.
 \item For every production $\rho \; = \;A\to\delta_2(B_1,\ldots,B_{i-1},\eta,B_{i+1},\ldots,B_k)$ in $P$, with $\eta\in\TT{N}{1}{0}$, the productions $A\to\delta_2(B_1,\ldots,B_{i-1},C_\rho,B_{i+1},\ldots,B_k)$ and $C_\rho\to \eta$ are in $P_1$.
 \item For every production $\rho \;=\; A(x)\to\delta_2(B_1,\ldots,B_{i-1},\eta,B_{i+1},\ldots,B_k)$ in $P$, with $\eta\in \lTT{N}{1}{1}$, the productions $A(x)\to\delta_2(B_1,\ldots,B_{i-1},C_\rho(x),B_{i+1},\ldots,B_k)$ and $C_\rho(x)\to \eta$ are in $P_1$.
\end{enumerate}

It is easy to see that $\lang(G_{1})=\lang(G)$. Now consider a production
\[\rho' \; = \; A\to \delta_{2}(B_{1},\ldots,B_{i-1},C_{\rho},B_{i+1},\ldots,B_{k})\]
in $P_{1}$. By Observations~\ref{obs:shape} and~\ref{obs:nonterminals}, $B_{j} \ne A$ for each $j\in[k]\setminus\{i\}$.  For this reason and since $C_{\rho}$ is a fresh nonterminal, $A$ does not occur in the right-hand side of $\rho'$.  

Thus, we can \emph{eliminate} the production \(\rho'\) from \(G_1\), as described in \cite[Def. 11]{Maletti2012}.  We construct an lm-cftg $\mathrm{Elim}(G_{1},\rho')$ as follows: for each production $s \to t$ in $P_{1}$ and each \(W \subseteq \{w \in \pos(t) \mid t(w) = A\}\), we construct a new production $s \to t'$ and insert it into \(P_1\). The new right-hand side $t'$ is obtained by substituting the right-hand side of \(\rho'\) for $A$ at each position in $W$. Then \(\rho'\) is removed from \(P_1\).  It was shown in \cite[Lemma~12]{Maletti2012} that $\lang(\mathrm{Elim}(G_{1},\rho'))=\lang(G_{1})$.
The same idea works for productions of the form $A(x)\to \delta_{2}(B_{1},\ldots,B_{i-1},C_{\rho}(x),B_{i+1},\ldots,B_{k})$ in $P_{1}$.

As an example, when we eliminate the production 
\(\rho' = C(x) \to \delta_2(A, B(x))\) in \(G_1\),
the production
\(A(x) \to \delta_1\bigl(B, C(D(x)), E\bigr)\) in \(G_1\)
results in \emph{two} new productions,
\[\rho_1 \; = \; A(x) \to \delta_1\bigl(B, C(D(x)), E\bigr) \qquad \text{and} \qquad \rho_2 \; = \; A(x) \to \delta_1\bigl(B, \delta_2(A, B(D(x))), E\bigr)\,\text{,} \]
and \(\rho'\) is discarded.

By applying this procedure successively for each production with a nonterminal from \(\tilde N\) on its left-hand side, we obtain in finitely many steps an equivalent lm-cftg $G_{2}$.  Note that \(G_2\) ``nearly'' has property \textit{(P)}: it may still contain productions which are not of the desired form.

In our example, if \(\rho'\) was the last production to be eliminated, then there is still the production \(\rho_1\) left, where \(\delta_2\) does not occur under \(\delta_1\).  However, it is easy to see that this production is \emph{useless:} after all, there are no productions left for the nonterminal \(C\).

This observation applies to all productions \(s \to t\) which are not of the desired form.  Therefore, \(\lang(G,t) = \emptyset\), and by a lemma of Rounds \cite[p.~113]{Rounds1970a}, we can just remove all these useless productions, resulting in the lm-cftg \(G_3 = (N_3, \Delta, S, P_3)\), which has property \textit{(P).}

\sepstars

We now use the same idea as in \cite{Arnold1980}.  As \(\delta_1\) and \(\delta_2\) appear right beneath each other in the productions of \(G_3\), they can just be replaced by \(\sigma\).

Formally, define a homomorphism \(\phi \colon \T_{N_3 \cup \Sigma}(X) \to \T_{N_3 \cup \Delta}(X)\) such that \(\phi(A) = A\) for each \(A \in N_3\) and \(\phi \vert_\Sigma = h\).  We construct an lm-cftg \(G' = (N_3, \Sigma, S, P')\) such that for each \(A \in N_3\) and \(t \in \TT{N_3 \cup \Sigma}{1}{1}\), we have
that \(A \to t\) is in \(P'\) if and only if \(A \to \phi(t)\) is in \(P_3\).

The formal proof that \(\lang(G') = h^{-1}(\lang(G))\) is omitted.
\end{proof}


\section{Conclusion}
In this work, we proved that the class of linear context-free tree languages is not closed under inverse linear tree homomorphisms.  However, the tree languages of linear monadic context-free tree grammars, which are employed in praxis under the pseudonym of tree-adjoining grammars, are closed under this operation.

In applications which require nonmonadicity and closure under inverse homomorphisms, it may prove beneficial to revisit the formalism of \emph{\(k\)-algebraic grammars,} i.e.\ context-free tree grammars over magmoids, where a nonterminal may derive a tuple of trees \cite[Chapter~V]{Arnold1978a}.  The class of languages defined by this type of grammar is indeed closed under inverse linear tree homomorphisms.

\bigskip
\noindent\textbf{Acknowledgement}\quad Last but foremost, we want to thank André Arnold for his help in this work.  In our email conversations, which we enjoyed very much, he showed us the flaws in our first proof attempts, and encouraged us to try on.  Moreover, the idea for the intermediate normal form of \(G\) in Lemma~\ref{lem:spine-grammar-three} is due to him, and he showed us how to significantly improve the presentation of the results in Sections~\ref{sec:dyck-words-sequences} and~\ref{sec:witness-langg-ne}.  

\bibliographystyle{abbrv}



\end{document}